\documentstyle[12pt]{article}

\newcommand{\om}{\omega}
\newcommand{\al}{\alpha}
\newcommand{\ep}{\epsilon}

\newcommand{\df}{\stackrel{\rm def}{=}}

\newcommand{\lb}{\lbrack}
\newcommand{\rb}{\rbrack}

\newcommand{\msc}[1]{\mbox{\scriptsize #1}}
\newcommand{\dsp}{\displaystyle}

\newcommand{\bc}{\mbox{{\bf C}}}

\newcommand{\bz}{\mbox{{\bf Z}}}

\newcommand{\bF}{\mbox{{\bf F}}}

\newcommand{\bsz}{\msc{{\bf Z}}}

\newcommand{\bsF}{\msc{{\bf F}}}

\newcommand{\da}{\dot{a}}

\newcommand{\db}{\dot{b}}

\newcommand{\cN}{{\cal N}}

\newcommand{\cS}{{\cal S}}

\newcommand{\cF}{{\cal F}}

\newcommand{\cH}{{\cal H}}

\newcommand{\ket}[1]{{|#1\rangle}}
\newcommand{\bra}[1]{{\langle#1|}}

\newcommand{\th}{{\theta}}

\newcommand{\tr}{\mbox{Tr}}

\newcommand{\NS}{\mbox{NS}}
\newcommand{\tNS}{\widetilde{\mbox{NS}}}
\newcommand{\R}{\mbox{R}}
\newcommand{\tR}{\widetilde{\mbox{R}}}
\newcommand{\sNS}{\msc{NS}}
\newcommand{\stNS}{\widetilde{\msc{NS}}}
\newcommand{\sR}{\msc{R}}
\newcommand{\stR}{\widetilde{\msc{R}}}

\newcommand{\hg}{\hat{\gamma}}
\newcommand{\hm}{\hat{\mu}}
\newcommand{\mbp}{\mbox{p}}
\newcommand{\msp}{\msc{p}}

\newcommand{\B}[1]{\mbox{\bf #1}}

\newcommand{\nn}{\nonumber\\}

\newcommand {\eqn}[1]{(\ref{#1})}

\makeatletter
\@addtoreset{equation}{section}
\def\theequation{\thesection.\arabic{equation}}
\makeatother

\begin{document}
\vskip 7mm

\begin{titlepage}
 \
 \renewcommand{\thefootnote}{\fnsymbol{footnote}}
 \font\csc=cmcsc10 scaled\magstep1
 {\baselineskip=14pt
 \rightline{
 \vbox{\hbox{hep-th/0209145}
       \hbox{UT-02-52}
       }}}

 \vfill
 \baselineskip=20pt
 \begin{center}
 \centerline{\Huge  Thermal Amplitudes} 
 \vskip 5mm 
 \centerline{\Huge   in DLCQ Superstrings on PP-Waves}

 \vskip 2.0 truecm
\noindent{\it \large Yuji Sugawara} \\
{\sf sugawara@hep-th.phys.s.u-tokyo.ac.jp}
\bigskip

 \vskip .6 truecm
 {\baselineskip=15pt
 {\it Department of Physics,  Faculty of Science, \\
  University of Tokyo \\
  Hongo 7-3-1, Bunkyo-ku, Tokyo 113-0033, Japan}
 }
 \vskip .4 truecm

 \end{center}

 \vfill
 \vskip 0.5 truecm

\begin{abstract}
\baselineskip 6.7mm

We calculate the thermal partition function of DLCQ superstring
on the maximally supersymmetric pp-wave background, which is 
realized  as the Penrose limit of orbifolded $AdS_5\times S^5$ 
and known to be dual to the  $\cN=2$ ``large'' quiver gauge theory 
as shown by S.~Mukhi, M.~Rangamani and E.~Verlinde, hep-th/0204147.
Making use of the path-integral technique, we derive the manifestly
modular invariant expression and show the equivalence with 
the free energy of second quantized free superstring 
on this background.  The ``virtual strings'' wound 
around  the temporal circle play essential roles for realizing 
the modular invariance and for a simple analysis on the Hagedorn temperature. 
We also calculate the thermal one-loop amplitudes of open strings 
under the various backgrounds of the supersymmetric time-like and 
Euclidean D-branes, and confirm the existence of correct open-closed 
string duality.
Furthermore, we extend these thermodynamical analysis to the 6-dimensional 
DLCQ pp-waves with general RR and NSNS flux. These superstring vacua 
are similarly derived from the supersymmetric (half SUSY) 
and non-supersymmetric orbifolds of $AdS_3 \times S^3  \times M^4 $ 
($M^4 =T^4 $ or $K3$) by the appropriate Penrose limits, 
giving rise to the SUSY enhancement as in the case of orbifolded 
$AdS_5 \times S^5$.

\end{abstract}

\vfill

\setcounter{footnote}{0}
\renewcommand{\thefootnote}{\arabic{footnote}}
\end{titlepage}
\baselineskip 18pt

\newpage
\section{Introduction}
\indent

String theories/M-theory on pp-wave backgrounds \cite{old pp}
have been recently studied with newer motivations. 
Among other things, it is remarkable that a new superstring vacuum
with the maximal SUSY has been discovered and derived from 
the $AdS_5 \times S^5$ background 
by the Penrose limits \cite{BFHP}. This string vacuum is
exactly quantizable in spite of the non-trivial RR-flux 
by the light-cone GS formalism \cite{Metsaev},
and provides a powerful tool to investigate the stringy nature 
of AdS/CFT correspondence beyond the supergravity approximation
\cite{BMN}.  

From the viewpoints of world-sheet theory,  
the light-cone string theories on such pp-wave backgrounds 
with the RR-flux exhibit a peculiar feature, namely, 
 {\em massive\/} world-sheets. This apparent lack 
of conformal symmetry seems to make it quite non-trivial 
to check whether these string vacua are really consistent,
for instance, to check the modular invariance, the open-closed string 
duality for the cylinder amplitudes (Cardy condition), and so on.
In particular, focusing on the problem of open-closed duality, 
a naive treatment would induce a difficulty, because we cannot 
take the light-cone gauge $X^+ \propto \tau$ at the same time
for both of the open and closed string channels.

Since the general pp-waves have a translational symmetry along a 
light-like direction, one can always consider the DLCQ 
(discrete light-cone quantized) string theories \cite{DLCQ} 
on these backgrounds. In this paper we shall study the DLCQ 
superstrings on the pp-waves with enhanced SUSY, and analyze 
the one-loop thermal amplitudes of closed strings and open strings
with supersymmetric D-brane backgrounds. Several motivations
for this study are in order:

Firstly, as explained in \cite{MRV} (see also \cite{ASJ2}), 
a nice realization of the DLCQ pp-wave with maximal SUSY is 
given by considering 
the Penrose limit of the orbifolded $AdS_5\times S^5$,
which is known to be dual to the $\cN=2$ quiver gauge theory
and has 16 supercharges. The Penrose limit is characterized 
by picking up a particular configuration of null-geodesic. 
If we choose it along the fixed point locus, 
it simply leads to the orbifolded pp-wave 
that has the same number of supercharges 
\cite{IKM,ASJ,orbifold pp}\footnote{The Penrose limits of 
several orbifolds of $AdS_5\times S^5$ and $AdS_5 \times T^{1,1}$ 
that are dual to the $\cN=1$ gauge theories
have been also studied in \cite{OhT}.}.  
On the other hand, if we place the null-geodesic away from
the fixed point locus, we obtain a smooth pp-wave 
with some compactification along longitudinal directions \cite{ASJ}.  
Moreover, the DLCQ limit is shown to correspond to
the ``large quiver limit'', which is a certain double scaling
limit considering the large $N$ and the large ``size'' of quiver diagram
(``deconstruction limit'' \cite{deconstruction}) at the same time 
\cite{MRV,ASJ2}.
In the latter case, which is of primary interest in this paper,
the space-time SUSY is maximally enhanced 
(32 supercharges), because the background has no orbifold singularity
and DLCQ does not break any supercharges. 
This is one of the well-known phenomena of SUSY enhancements
under the several Penrose limits discussed by many authors 
(see, for example, \cite{IKM,GNS}). 

Secondly, the DLCQ string theory is known to have
effectively discretized  moduli  of world-sheet. 
Therefore, it seems comparably easy to observe
how the modular invariance and the open-closed duality  
are realized. We will later demonstrate how these consistencies  
are established in the framework of thermal string theory.

Thirdly, our thermodynamical analysis on  DLCQ pp-waves
may shed new light on the several 
attempts for the Matrix string theories \cite{MST}
describing pp-wave backgrounds \cite{Verlinde,MST pp}. 
In fact, in the case of flat background, it is known that 
the free energy of Matrix string theory coincides 
with that of the second quantized DLCQ superstring \cite{Semenoff}.

Fourthly, we would like to also mention on the models of 
4-dimensional NSNS pp-wave with enhanced SUSY constructed in \cite{HS2}.
These superstring vacua are defined based on the super Nappi-Witten model 
\cite{NW,KK,Sfetsos} and arbitrary rational $\cN=2$ SCFT with $c=9$, 
being orbifolded by the GSO projection like the Gepner models
\cite{Gepner}.
We point out  that these models have the light-cone momentum 
discretized by the GSO condition just mentioned, and hence  
show the feature quite reminiscent of the DLCQ pp-wave.

~

This paper is organized as follows:

In section 2 we calculate the thermal partition function of IIB
superstring on the 10-dimensional DLCQ pp-wave mentioned above.
By making use of the path-integral technique we derive the manifestly
modular invariant expression, and further confirm that it 
actually coincides  with the free energy of second quantized
string theory calculated by the operator formalism defined over the physical 
Hilbert space. The existence of the ``virtual strings''  
wound around the temporal circle (or, we call it ``thermal circle'')
is quite important for the modular invariance 
and a simple analysis on the Hagedorn temperature \cite{Hagedorn}. 
In section 3 we analyze the thermal one-loop amplitudes 
of open strings under the supersymmetric backgrounds of 
the time-like and Euclidean D-branes. We especially focus on the problem
how the open-closed string duality should be understood in the context
of thermal string theory on the DLCQ pp-waves. 
The virtual string sectors again play an essential role.
In section 4, we extend our analysis to the cases of the 6-dimensional 
DLCQ pp-waves with the enhanced SUSY, which are similarly derived from 
the non-SUSY and half-SUSY orbifolds of $AdS_3\times S^3$ backgrounds.
We give a summary and discussions in section 5.

~

We should finally comment on some recent works related to this paper. 
One-loop amplitudes for non-thermal, non-DLCQ string theory 
on the 10-dimensional pp-wave have been analyzed  for open string
in \cite{BGG},  and for closed string in \cite{Takayanagi}.
The thermal partition function for closed string in the non-DLCQ
model has been calculated in \cite{PV,GSS}.

~


\section{Thermal Partition Function of DLCQ Superstring on 10-dimensional
PP-Wave}

\subsection{Short Review of the Light-cone GS Superstring on Maximally
Supersymmetric PP-Wave}
\indent

It is familiar that type IIB string on
the maximally supersymmetric pp-wave background
is canonically quantized in the light-cone GS formalism \cite{Metsaev}. 
We shall start with a brief review of it mainly to prepare the notations.

We introduce the bosonic string coordinates
$\dsp X^{\pm} \equiv \frac{1}{\sqrt{2}}(X^9\pm X^0)$, $X^I$
($I=1,\ldots,8$), and the GS fermions $\theta^A$, $\tilde{\theta}^A$
which are 10-dimensional Majorana-Weyl spinors having the same
chirality. The relevant pp-wave geometry is expressed  as 
\begin{eqnarray}
ds^2 = 2dX^+dX^- - \mu^2 (X^I)^2 (dX^+)^2+ 
(dX^I)^2~,
\label{pp-wave}
\end{eqnarray}
with the RR 5-form flux
\begin{equation}
F_{+1234}=F_{+5678}\sim \mu~.
\label{5-form flux}
\end{equation}
The light-cone gauge is defined by
\begin{eqnarray}
X^+= \al' p^+\tau~,~~~\Gamma^+\theta=\Gamma^+\tilde{\theta}=0~.
\label{lc gauge}
\end{eqnarray}
We write the remaining 8 component spinors 
as $S^a$, $\tilde{S}^a$ (with a  conventional rescaling) 
composing the spinor representation ${\bf 8}_s$ of 
$SO(8)$ respectively.
It is convenient to introduce the chiral representation of $SO(8)$ gamma 
matrices as
\begin{eqnarray}
\hat{\gamma}^I=
\left(
\begin{array}{cc}
 O & \gamma_{a\db} \\
 \bar{\gamma}_{\da b} & O
\end{array}
\right)~,  ~~~ \{\hat{\gamma}^I,\, \hat{\gamma}^J\} = 2\delta^{IJ}~.
\end{eqnarray}
The $8\times 8$ matrices $\gamma^I_{a\db}$, $\bar{\gamma}^I_{\da b}$
clearly satisfy 
\begin{eqnarray}
 \gamma^I_{a\db}\bar{\gamma}^J_{\db c} 
+ \gamma^J_{a\db} \bar{\gamma}^I_{\db c} = 2\delta^{IJ}\delta_{ac}~,~~~
\bar{\gamma}^I_{\da b} \gamma^J_{b\dot{c}}
+ \bar{\gamma}^J_{\da b} \gamma^I_{b\dot{c}}= 
2\delta^{IJ}\delta_{\da \dot{c}}~.
\end{eqnarray}
We can assume all $\hat{\gamma}^I$ are real symmetric, resulting that
$(\gamma^I)^T= \bar{\gamma}^I$.  
The light-cone gauge action is now  written as
\begin{eqnarray}
S &=& \frac{1}{4\pi \al'}\int d^2\sigma \, 
\left(\partial_+ X^I \partial_- X^I - m^2 (X^I)^2 \right) ~ \nn
&& ~~~~~ +\frac{i}{2\pi} \int d^2 \sigma\,
\left(S^a \partial_+ S^a + 
\tilde{S}^a\partial_-\tilde{S}^a -2mS^a\Pi_{ab}\tilde{S}^b
\right)~,
\label{lc action}
\end{eqnarray}
where we set $\Pi = \gamma^1\bar{\gamma}^2\gamma^3\bar{\gamma}^4$.
It is easy to see $\Pi^T = \Pi$, $\Pi^2=1$. 
We denote $\partial_{\pm}=\partial_{\tau}\pm \partial_{\sigma}$ as usual.
The mass parameter $m$ is defined as $m=   \mu \al' p^+$.
The existence of mass terms breaks the conformal symmetry  on
world-sheet,  and moreover  the fermionic mass term breaks 
the  global symmetry down to $SO(4)\times SO(4)$.

The equations of motion are given by
\begin{eqnarray}
&& \partial_+\partial_- X^I + m^2 X^I = 0 ~,\nn
&& \partial_+S^a-m\Pi_{ab}\tilde{S}^b=0~,\nn
&& \partial_-\tilde{S}^a+m\Pi_{ab}S^b=0~.
\label{eom}
\end{eqnarray}
The solutions with periodic boundary conditions have the 
following mode expansions;
\begin{eqnarray}
&&X^I(\tau,\sigma)= x_0^I \cos(m\tau)+\frac{\al'}{m}p^I_0\sin(m\tau)
+\sqrt{\frac{\al'}{2}}\sum_{n\neq 0}\frac{1}{\sqrt{\om_n}}
\left(a^I_n e^{-i(\om_n\tau-n\sigma)}+
a^{\dag I}_n e^{i(\om_n\tau-n\sigma)}
\right)~, \nn
&& \\
&&P^I(\tau,\sigma)\equiv \frac{1}{2\pi \al'}\partial_{\tau}X^I \nn
&& ~~~ = -\frac{m}{2\pi \al'}x_0^I \sin(m\tau)+\frac{1}{2\pi}p^I_0\cos(m\tau)
-\frac{i}{2\sqrt{2 \al'}\pi}\sum_{n\neq 0}\sqrt{\om_n}
\left(a^I_n e^{-i(\om_n\tau-n\sigma)}-
a^{\dag I}_n e^{i(\om_n\tau-n\sigma)}
\right)~,  \nn
&& \\
&& S^a(\tau,\sigma)=S_0^a \cos(m\tau)+\Pi_{ab}\tilde{S}^b \sin(m\tau)\nn
&& ~~~ + \sum_{n>0}\, c_n\left\lb S_n^a e^{-i(\om_n\tau-n\sigma)}
+ S_n^{\dag a} e^{i(\om_n\tau-n\sigma)} 
+i\frac{\om_n-n}{m} \Pi_{ab}\left(
S_{-n}^b e^{-i(\om_n\tau+n\sigma)}-S_{-n}^{\dag b} 
e^{i(\om_n\tau+n\sigma)} 
\right)   \right\rb ~, \nn
&& \\
&& \tilde{S}^a(\tau,\sigma)
=\tilde{S}_0^a \cos(m\tau)- \Pi_{ab}S^b \sin(m\tau)\nn
&& ~~~ + \sum_{n>0}\, c_n\left\lb S_{-n}^a e^{-i(\om_n\tau+n\sigma)}
+ S_{-n}^{\dag a} e^{i(\om_n\tau+n\sigma)} 
-i\frac{\om_n-n}{m} \Pi_{ab}\left(
S_{n}^b e^{-i(\om_n\tau-n\sigma)}-S_{n}^{\dag b} e^{i(\om_n\tau-n\sigma)} 
\right)   \right\rb ~, \nn
\end{eqnarray}
where we set
\begin{equation}
\om_n \equiv \sqrt{m^2+n^2}~, ~~~ c_n\equiv 
\frac{1}{\sqrt{1+\left(\frac{\om_n-n}{m}\right)^2}}~.
\end{equation}
The modifications of mode expansions  for more general boundary conditions
are quite easy and we do not write them explicitly.

The canonical quantization gives the standard (anti-)commutation 
relations of harmonic oscillators\footnote
   {We are here taking the convention such that the modes $n>0$ 
    correspond to the left-mover and $n<0$ to the right-mover 
    respectively under the conformal limit $m~\rightarrow~0$ according to
    \cite{BMN}.};
\begin{eqnarray}
&&\lb a^I_m,\, a^{\dag J}_n\rb = \delta^{I,J}\delta_{m,n}~, ~~~
\lb a^I_m,\, a^J_n \rb = \lb a^{\dag I}_m,\, a^{\dag J}_n \rb =0~, \\
&&\lb x^I_0, \, p^J_0 \rb = i \delta^{I,J}~, \\
&&\{ S^a_m,\, S^{\dag b}_n\} = \delta^{a,b}\delta_{m,n}~, ~~~
\{S^a_m,\,S^b_n\}=\{S^{\dag a}_m,S^{\dag b}_n\}=0~, \\
&&\{S^a_0, \, S^b_0\}=\delta^{a,b}~, 
~~~ \{\tilde{S}^a_0, \, \tilde{S}^b_0\}=\delta^{a,b}~,~~~
\{S^a_0, \, \tilde{S}^b_0\}=0~.
\end{eqnarray} 
We also introduce the next notations for the zero-mode oscillators
in order to diagonalize the zero-mode part of Hamiltonian;
\begin{eqnarray}
&& a^I_0\equiv \frac{1}{\sqrt{2m\al'}}(mx^I_0+i\al'p^I_0)~,~~~
a^{\dag I}_0\equiv \frac{1}{\sqrt{2m\al'}}(mx^I_0-i\al'p^I_0)~,\\
&&S^a_{\pm}\equiv \frac{1}{2}(1\pm \Pi)_{ab}
\frac{1}{\sqrt{2}}(S^b_0\pm i \tilde{S}^b_0)~,~~~
S^{\dag a}_{\pm}\equiv \frac{1}{2}(1\pm \Pi)_{ab}
\frac{1}{\sqrt{2}}(S^b_0\mp i \tilde{S}^b_0)~,
\end{eqnarray}
which satisfy
\begin{eqnarray}
&&\lb a^I_0,\, a^{\dag J}_0 \rb=\delta^{I,J}~, ~~~
\lb a^I_0,\, a^J_0 \rb = \lb a^{\dag I}_0,\, a^{\dag J}_0 \rb =0~,\\
&&\{S_{\pm}^a,\,S_{\pm}^b\}=\delta^{a,b}~,~~~
\{S_{\pm}^a,\,S_{\mp}^b\}=0~.
\end{eqnarray}
The light-cone Hamiltonian is calculated by the Virasoro constraints;
\begin{equation}
H_{\msc{l.c.}} \equiv -p^- = \frac{1}{\al' p^+} 
\sum_{n\in \bsz} \om_n N_n~ +a(p^+)^{(b)}+a(p^+)^{(f)},
\label{Hlc}
\end{equation}
where $N_n$ denotes the mode counting operators at the level $n$;
\begin{eqnarray}
N_n&=& a^{\dag I}_n a^I_n + S^{\dag a}_nS^a_n ~, ~~~ (n\neq 0) \nn
N_0&=& a^{\dag I}_0 a^I_0+ iS^a_0\Pi_{ab}\tilde{S}^b_0 +4 \nn
&\equiv& a^{\dag I}_0 a^I_0+ S^{\dag a}_+S^a_+
+S^{\dag a}_-S^a_- ~.
\end{eqnarray}
$a(p^+)^{(b)}$, $a(p^+)^{(f)}$ 
are the normal order constants for bosonic and fermionic
sectors respectively which may non-trivially depend on
$p^+$. In the present set up they should 
totally cancel because the bosonic and fermionic 
coordinates satisfy the same boundary condition.
We will face more non-trivial situations in 
which they have different boundary conditions and the cancellation fails.  
We will separately fix them later in order to calculate 
the thermal amplitudes.   

The Fock vacuum $\ket{0;p^+}$ 
is characterized in the standard manner;
\begin{equation}
a^I_n\ket{0;p^+}=0~({}^{\forall}I,n)~,~~~
S^a_n\ket{0;p^+}=0~({}^{\forall}a,{}^{\forall}n\neq 0)~,~~~
S^a_{\pm} \ket{0;p^+}=0~({}^{\forall}a)~.
\end{equation}

Since we are now interested in  the DLCQ string theory \cite{DLCQ}
$X^-\sim X^- + 2\pi R_-$, the light-cone momentum $p^+$
should be quantized as
\begin{equation}
p^+ = \frac{p}{R_-}~,~~~ (p \in \bz_{>0})~.
\label{dlcq momentum}
\end{equation}
The Virasoro constraints provide the level matching condition
\begin{equation}
\sum_{n\in \bsz}n N_n = pk ~, ~~~({}^{\forall} k \in \bz)~,
\label{level matching}
\end{equation}
for the each winding sector 
$\dsp \int_0^{2\pi} d\sigma \partial_{\sigma} X^- = 2\pi kR_-$.

~

\subsection{Transverse Partition Functions}
\indent

Now, we are ready to calculate the toroidal  partition function.
We first focus on the transverse sector. We so fix the mass parameter
$m$ for the time being.
According to the standard treatment, we move to the Euclidean
world-sheet by the Wick rotation $\tau=i \tau_E$, and set
\begin{equation}
z=i\tau_E-\sigma~,~~~
\bar{z}=-i\tau_E-\sigma ~.
\end{equation}
resulting the replacement $\dsp \partial_+~\rightarrow ~ 
-2 \partial_{\bar{z}} $, 
$\partial_-~\rightarrow~ 2\partial_z$.
We also introduce the next parameterizations
$z=\xi_1+\tau\xi_2$, 
$\bar{z}=\xi_1+\bar{\tau}\xi_2$, 
where $\tau=\tau_1+i\tau_2$ $(\tau_2>0)$
denotes the modulus parameter of world-sheet torus.
The next formulas are often useful for calculations;
\begin{eqnarray}
 \partial_z=\frac{i}{2\tau_2}(\bar{\tau}\partial_{\xi_1}-\partial_{\xi_2})~,
~~~
 \partial_{\bar{z}}
=-\frac{i}{2\tau_2}(\tau\partial_{\xi_1}-\partial_{\xi_2})~,~~~
d^2z= \tau_2 d\xi_1d\xi_2~,
\label{z xi}
\end{eqnarray} 

The transverse partition function is  
calculated in the way parallel to  
the standard conformal field theory;
\begin{equation}
Z^{\msc{tr}}(\tau, \bar{\tau};m) =\tr \left\lb (-1)^{\bsF}
\, e^{-2\pi \tau_2 H + 2\pi i \tau_1 P}\right\rb~,
\label{Ztr}
\end{equation} 
where $H\equiv \al'p^+H_{\msc{l.c.}}$ is the world-sheet Hamiltonian
and $\dsp P\equiv \sum_{n\in\bsz}n N_n$ is the world-sheet momentum 
operator. In the conformal limit we of course obtain 
$H=L_0+\tilde{L}_0-1$, $P=L_0-\tilde{L}_0$. $\bF$ denotes the space-time
fermion number  and the insertion of $(-1)^{\bsF}$ is necessary to realize 
the periodic boundary condition for the GS fermions.

Because our transverse Hilbert space is a free Fock space, the calculation 
of trace is quite easy except for the evaluation of zero-point energy 
(or, the normal order constant) with an appropriate  regularization.  
Let us first pick up one {\em complex\/} boson. 
According to \cite{BGG,Takayanagi,GSS}, we shall 
evaluate the regularized zero-point energy as the Casimir energy,
which is defined by subtracting the divergent contribution free to
the boundary condition. 
Since the zero-point energy for each of harmonic oscillators 
$a^I_n$, $a^{\dag I}_n$ is equal to 
$\om_n/2$, we can explicitly calculate it as (for the chiral part) 
\begin{eqnarray}
\Delta(m)& \df& 
\frac{1}{2}\left(\sum_{n\in\bsz}\sqrt{m^2+n^2}  - 
\int_{-\infty}^{\infty}dk\, \sqrt{m^2+k^2} \right) \nn
&=& 
\frac{1}{2}\sum_{n\in\bsz}\frac{1}{\Gamma(-1/2)}\int_0^{\infty}dt\,
t^{-3/2}e^{-t(m^2+n^2)} -\int_{-\infty}^{\infty}dk\,
\frac{1}{\Gamma(-1/2)}\int_0^{\infty}dt\,
t^{-3/2}e^{-t(m^2+k^2)}   \nn
&=& -\frac{1}{4}\sum_{n\neq 0} \int_0^{\infty}dt\, t^{-2} 
e^{-tm^2-\frac{\pi^2n^2}{t}} \nn
&=& -\frac{1}{2\pi^2} \sum_{n=1}^{\infty} \int_0^{\infty}ds\,
e^{-sn^2-\frac{\pi^2m^2}{s}} ,
\label{Delta m}
\end{eqnarray}
where we have used the Poisson resummation formula to derive the third line. 
This actually converges  by the evaluation
\begin{equation}
\sum_{n=1}^{\infty}\left|\int_0^{\infty}ds\,
e^{-sn^2-\frac{\pi^2m^2}{s}}\right| \leq
\sum_{n=1}^{\infty}\left|\int_0^{\infty}ds\,
e^{-sn^2}\right| = \sum_{n=1}^{\infty} \frac{1}{n^2}=\zeta(2)=\frac{\pi^2}{6}~.
\end{equation}
%
The  partition function for one complex boson now becomes
\begin{eqnarray}
Z_{\msc{boson}}(\tau,\bar{\tau};m)
=\frac{1}{e^{4\pi \tau_2 \Delta(m)} \, \prod_{n\in\bsz}
\left(1-e^{-2\pi \tau_2 \sqrt{m^2+n^2}+2\pi i\tau_1 n} \right)^2}~.
\label{part boson}
\end{eqnarray}
The regularized zero-point energy \eqn{Delta m}
is justified as follows: Firstly, it has the correct $m~\rightarrow~0$
limit 
\begin{equation}
\lim_{m\rightarrow 0} \Delta(m) = -\frac{1}{2\pi^2}
\sum_{n=1}^{\infty} \frac{1}{n^2}=-\frac{1}{12} 
\left(\equiv 2\times (-\frac{1}{24})\right)~. 
\end{equation}
Moreover, the partition function \eqn{part boson} 
possesses the correct modular properties based on the definition \eqn{Delta m},
as is proved in \cite{BGG,Takayanagi};
\begin{eqnarray}
Z_{\msc{boson}}(\tau+1,\bar{\tau}+1;m)&=&
Z_{\msc{boson}}(\tau,\bar{\tau};m)~, \nn
Z_{\msc{boson}}(-1/\tau,-1/\bar{\tau};m|\tau|)&=&
Z_{\msc{boson}}(\tau,\bar{\tau};m)~.
\label{modular 1}
\end{eqnarray}
The reason why these modular properties should be correct  
is most naively explained as follows: The modular transformations 
preserve the complex  structure of torus,  but the  $S$-transformation
$\tau~\rightarrow~-1/\tau$ changes the area as 
$\dsp A~\rightarrow~\frac{1}{|\tau|^2}A$. We now have a unique 
mass parameter $m$ that breaks conformal invariance 
and has the dimension $\lb\mbox{length}\rb^{-1}$. 
Therefore, the physical contents should not be changed, if the 
$S$-transformation is  compensated by 
the scale transformation $m~\rightarrow~|\tau|m$.
More rigorous understanding of the modular properties 
is achieved  by the path-integral  as discussed in \cite{PV}.
The Gaussian path-integral is easily evaluated and gives the determinant
of the Klein-Gordon operator $-4\partial_z\partial_{\bar{z}}+m^2$.
Working with  the coordinates $z=\xi_1+\tau\xi_2$, 
$\bar{z}=\xi_1+\bar{\tau}\xi_2$ and making use of \eqn{z xi},
we can show
\begin{eqnarray}
Z_{\msc{boson}}(\tau,\bar{\tau};m) 
\sim \left\lb  \prod_{n_1,n_2 \in \bsz} \,
\tau_2 \left(\frac{1}{\tau_2^2}|n_1-n_2\tau|^2
  + m^2\right)\right\rb^{-1}~,
\label{path integral boson}
\end{eqnarray}
up to a divergent factor that is  independent of $\tau$, 
$\bar{\tau}$, $m$ and  should be regulated.
It manifestly reproduces the modular transformation formulas
\eqn{modular 1}.
The zero-point energy \eqn{Delta m} is justified by these reasons.

For later convenience we further introduce the ``massive
theta function'' according to \cite{Takayanagi};
\begin{eqnarray}
&& \Theta_{(a,b)}(\tau,\bar{\tau};m) \df
e^{4\pi\tau_2 \Delta(m;a)}\prod_{n\in\bsz}\left|
 1-e^{-2\pi \tau_2 \sqrt{m^2+(n+a)^2} +2\pi i\tau_1(n+a)+2\pi i b}
\right|^2~.
\label{massive theta}
\end{eqnarray}
where $a,b$ are arbitrary real parameters. The zero-point 
energy $\Delta(m;a)$ is similarly defined by
\begin{eqnarray}
\Delta(m;a) &\df& \frac{1}{2}\sum_{n\in\bsz}\sqrt{m^2+(n+a)^2}
-\frac{1}{2}\int_{-\infty}^{\infty} dk\, \sqrt{m^2+k^2} \nn 
&=&
 -\frac{1}{2\pi^2} \sum_{n=1}^{\infty}
\int_{0}^{\infty}ds\,
e^{-sn^2-\frac{\pi^2m^2}{s}} \cos(2\pi n a)~.
\label{Delta m a} 
\end{eqnarray}
The function $\Theta_{(a,b)}(\tau,\bar{\tau};m)$ describes 
various twisted boundary conditions
characterized by $a$ and $b$. Namely, it is easy to see
that the partition function for the $d$-components complex
massive boson (non-chiral fermion) with the boundary conditions
$\phi(z+2\pi, \bar{z}+2\pi) = e^{-2\pi i a}\phi(z,\bar{z})$ \footnote
     {Recall that $z\,\rightarrow\,z+2\pi$, $\bar{z}\,\rightarrow\,
     \bar{z}+2\pi$ corresponds to $\sigma\,\rightarrow \, \sigma -2\pi$.
      The extra minus sign defining this twisted boundary condition
       is due to  this fact.}, 
$\phi(z+2\pi\tau, \bar{z}+2\pi\bar{\tau}) = e^{2\pi i b}\phi(z,\bar{z})$,
is calculated  as  
\begin{eqnarray}
Z(\tau,\bar{\tau};m)& =&
\tr\left\lb (-1)^{\bsF}\, e^{-2\pi \tau_2 H + 2\pi i \tau_1 \hat{P}
+2\pi i b \hat{h} 
}\right\rb   \nn
&=& \Theta_{(a,b)}(\tau,\bar{\tau};m)^{-\ep d}~, 
\label{part m t boson}
\end{eqnarray}
where $\ep = +1 $ for the boson and $\ep=-1$ for the fermion.  
In this expression we introduced the momentum operator for the 
twisted fields
\begin{eqnarray}
\hat{P} = \sum_n\left((n+a)N^{(+)}_n + (n-a)N^{(-)}_n\right)~,
\label{hat P 0}
\end{eqnarray}
and the ``helicity operator''
\begin{eqnarray}
\hat{h} = \sum_n \left(N^{(+)}_n-N^{(-)}_n\right)~,
\label{hat h}
\end{eqnarray}
where $N^{(+)}_n$, $N^{(-)}_n$ express the mode counting operators
associated to the Fourier modes 
$e^{\pm i(n+a)\sigma}$, $e^{\pm i(n-a)\sigma}$ respectively.

$\Theta_{(a,b)}(\tau,\bar{\tau};m)$ has the following 
modular properties
\begin{eqnarray}
&& \Theta_{(a,b)}(\tau+1,\bar{\tau}+1;m) 
= \Theta_{(a,b+a)}(\tau,\bar{\tau};m)~, \nn
&& \Theta_{(a,b)}(-1/\tau,-1/\bar{\tau};m|\tau|) 
= \Theta_{(b,-a)}(\tau,\bar{\tau};m)~.
\label{modular 2}
\end{eqnarray}
We can also show that
\begin{eqnarray}
&& \Theta_{(a,b)}(\tau,\bar{\tau};m)=\Theta_{(-a,-b)}(\tau,\bar{\tau};m)
=\Theta_{(a+r,b+s)}(\tau,\bar{\tau};m)~, ~~~({}^{\forall}r,s \in \bz)~, \\
&& \lim_{m\rightarrow 0} \Theta_{(a,b)}(\tau,\bar{\tau};m)
= e^{-2\pi \tau_2 a^2}\,\left|
\frac{\th_1(\tau,a\tau+b)}{\eta(\tau)}
\right|^2~.
\end{eqnarray}

In our present  problem the transverse partition function 
\eqn{Ztr} is calculated as
\begin{eqnarray}
&& Z^{\msc{tr}}(\tau, \bar{\tau};m) =\tr \left\lb (-1)^{\bsF}
\, e^{-2\pi \tau_2 H + 2\pi i \tau_1 P}\right\rb = 
\frac{\Theta_{(0,0)}(\tau,\bar{\tau};m)^4}
{\Theta_{(0,0)}(\tau,\bar{\tau};m)^4} \equiv 1~.
\label{Ztr 2}
\end{eqnarray}
We also obtain for the twisted boundary conditions for the fermionic
coordinates
\begin{eqnarray}
&& \tr \left\lb 
 e^{-2\pi \tau_2 H + 2\pi i \tau_1 P}\right\rb = 
\frac{\Theta_{(0,1/2)}(\tau,\bar{\tau};m)^4}
{\Theta_{(0,0)}(\tau,\bar{\tau};m)^4} ~, \nn
&& \tr_{\cH^{(t)}} \left\lb 
 (-1)^{\bsF}\, e^{-2\pi \tau_2 H + 2\pi i \tau_1 P}\right\rb = 
\frac{\Theta_{(1/2,0)}(\tau,\bar{\tau};m)^4}
{\Theta_{(0,0)}(\tau,\bar{\tau};m)^4} ~, \nn
&&\tr_{\cH^{(t)}} \left\lb 
e^{-2\pi \tau_2 H + 2\pi i \tau_1 P}\right\rb = 
\frac{\Theta_{(1/2,1/2)}(\tau,\bar{\tau};m)^4}
{\Theta_{(0,0)}(\tau,\bar{\tau};m)^4}~,
\label{Ztr 3}
\end{eqnarray}
where $\tr_{\cH^{(t)}}$ means the trace over the Hilbert
space of the anti-periodic
GS fermions $S^a(z+2\pi, \bar{z}+2\pi) = -S^a(z, \bar{z})$,
$\tilde{S}^a(z+2\pi, \bar{z}+2\pi) = -\tilde{S}^a(z, \bar{z})$.

We finally comment on the result \eqn{Ztr 2}. In contrast to the flat
background the partition function \eqn{Ztr 2} 
does not vanish although we have maximal SUSY. 
This aspect is understood as follows.
We have the 16 kinematical supercharges,
which are essentially the zero-modes of GS fermions,
and also the 16 dynamical supercharges including the higher level oscillators.
The latter commutes with the light-cone Hamiltonian (and hence $H$, too),
but the former does  not, contrary to the flat case in which
all the supercharges commute with Hamiltonian. 
In this situation, since the GS partition function \eqn{Ztr}  
is a Witten index by definition, we could obtain non-zero contributions from
the BPS states, which are annihilated  by the 
dynamical supercharges, or equivalently, have the vanishing light-cone
energy. In fact, we now have a unique BPS state for each $p^+$, 
namely, the Fock vacuum $\ket{0;p^+}$ itself, 
and hence  the Witten index should be equal to 1.
This property of partition function 
is already mentioned in \cite{Takayanagi} and the similar feature 
in  the open string one-loop amplitudes is also pointed out in
\cite{BGG}.

One might still ask: {\em 
Does it mean the non-zero cosmological constant?}  
{\em Cannot we  continuously take the flat limit $\mu \rightarrow 0$
in the level of partition function?}

These are really  not the cases. Recall that the cosmological 
constant should be identified with the vacuum energy {\em density\/}, 
and we have a divergent volume since the pp-wave background 
is non-compact. In the case of flat background the partition function
includes the volume factor due to the zero-modes of bosonic 
coordinates. However, such volume factor does not appear in the present
case, since the zero-modes $a^I_0$, $a^{\dag I}_0$, 
$\ldots$ possess  non-zero energy due to mass $m$.  
In this sense the vacuum energy density surely behaves continuously
under the $\mu ~\rightarrow ~0$ limit (namely, remains zero).  
In this limit  the transverse volume factor $V_8$  appears from the bosonic
contribution
\begin{equation}
\lim_{m\rightarrow 0} \frac{1}{\Theta_{(0,0)}(\tau,\bar{\tau};m)^4}
\sim V_8\times \frac{1}{\tau_2^4} \frac{1}{|\eta(\tau)|^{16}}~,
\end{equation}
which cancels the denominator to define the vacuum energy density.
On the other hand, 
the fermionic contribution precisely cancels with  each other
this time, resulting also the vanishing energy density.

~

\subsection{Thermal Partition Function}
\indent

Now, let us proceed to our main subject, the calculation of  
the thermal partition function for the DLCQ string on pp-wave background.
The thermal string theory is defined as the target space with  the 
compactified Euclidean time with the circumference equal to 
the inverse temperature $\beta$. 

According to \cite{Semenoff},
we shall employ the path-integral technique for the longitudinal sector. 
The advantage to do so is that we can most transparently obtain  
the modular invariant expression of partition function.
First of all, we consider the thermal DLCQ string 
in the flat background as a warm-up.
In the Wick rotated space-time 
$\dsp X^{\pm}\equiv \frac{1}{\sqrt{2}}(X^9\pm i X_E^0)$,
the DLCQ string theory ($X^-\sim X^-+2\pi R_-$) is 
described by the complex identification
\begin{equation}
X^0_E\sim X^0_E+\sqrt{2}\pi R_- i~,~~~ X^9\sim X^9+\sqrt{2}\pi R_-~,
\label{E DLCQ}
\end{equation}
and the thermal compactification is defined as
\begin{equation}
X^0_E\sim X^0_E+\beta~,
\end{equation}
where $\beta$ denotes the inverse temperature.
The complex identification \eqn{E DLCQ} may sound peculiar, 
since it makes the world-sheet action complex\footnote
   {As well as the Euclidean DLCQ string theory, we obtain
    the complex world-sheet actions  
    for the Wick rotated theories of general pp-waves 
    (including the DLCQ pp-waves, of course). Such string models
    seem to be ill-defined as canonically quantized theories 
    based on these complex actions themselves. 
    However, if taking  the thermal compactification at the 
    same time, they provide an useful machinery to derive 
    modular invariant amplitudes by the path-integration. 
    Modestly speaking, to adopt the path-integral approach based on 
    these complex world-sheet actions at least has a physical meaning as
    a conventional method to calculate the free energies of 
    the original pp-wave string theories with Lorentzian signature.  
    We will later confirm the equivalence with the operator
    formalism, which justifies this approach.}. 
Nevertheless, it has been proved  in  \cite{Semenoff} that for
the flat background it gives the results equivalent with
the operator formalism defined with respect to the original theory
of Lorentzian signature, in which only the physical states 
appear in the calculation. The path-integral approach presented here
is justified by this fact.

When calculating the Polyakov path-integral,  
the contributions from the various topological sectors  
are most important;
\begin{eqnarray}
 X^+(z+2\pi,\bar{z}+2\pi)&=& X^+(z,\bar{z})+\frac{i\beta}{\sqrt{2}}w ~,\nn
 X^+(z+2\pi\tau,\bar{z}+2\pi\bar{\tau})
&=& X^+(z,\bar{z})+\frac{i\beta}{\sqrt{2}}n ~,\nn
X^-(z+2\pi,\bar{z}+2\pi)&=& X^-(z,\bar{z})-\frac{i\beta}{\sqrt{2}}w 
+2\pi R_- r~,\nn
X^-(z+2\pi\tau,\bar{z}+2\pi\bar{\tau})
&=& X^-(z,\bar{z})-\frac{i\beta}{\sqrt{2}}n +2\pi R_- s~, \nn
&& \hspace{1in} (w,n,r,s \in \bz)~.
\label{instanton}
\end{eqnarray}
The ``instanton'' solutions are  defined as those which linearly depend on
the world-sheet coordinates and are constrained by these boundary 
conditions.
The instanton action is evaluated with the helps of \eqn{z xi} as
\begin{eqnarray}
S_{\msc{inst}}(w,n,r,s)
=\frac{\beta^2|w\tau-n|^2}{4\pi\al'\tau_2}
+ 2\pi i \frac{\nu}{\tau_2}\left\{|\tau|^2wr-\tau_1(ws+nr)+ns\right\}~,
\label{inst action 1}
\end{eqnarray}
where we set $\dsp \nu \equiv \frac{\sqrt{2}\beta R_-}{4\pi \al'}$.
The longitudinal path-integral is evaluated as
\begin{eqnarray}
\frac{V_{\msc{l.c.}}}{4\pi^2 \al' \tau_2} \times \sum_{w,n,r,s}\,
e^{-S_{\msc{inst}}(w,n,r,s)} = \frac{\nu}{\tau_2}
\sum_{w,n,r,s}\, e^{-S_{\msc{inst}}(w,n,r,s)}~,
\end{eqnarray}
where the $V_{\msc{l.c.}}\equiv \sqrt{2}\pi \beta R_-$ is the volume of 
longitudinal directions. The prefactor 
$V_{\msc{l.c.}}/(4\pi^2 \al' \tau_2)$ is given by  
integrating out the fluctuations around instantons, also taking account of
the FP determinant in the standard manner. In other words it 
corresponds to the factor derived from the Gaussian integral 
of the zero-mode momenta in the Hamiltonian formalism.

Another non-trivial point is the boundary conditions of fermionic 
coordinates along the thermal circle \cite{AW}.
The winding numbers $w$, $n$ are respectively the
spatial and temporal ones.
We should hence choose the boundary conditions for GS fermions as 
\begin{eqnarray}
&& S^a(z+2\pi \ep_1+2\pi\ep_2\tau) = (-1)^{\ep_1 w+\ep_2 n} S^a(z)~,~~~
\tilde{S}^a(\bar{z}+2\pi \ep_1+2\pi\ep_2\bar{\tau}) 
= (-1)^{\ep_1 w+\ep_2 n} \tilde{S}^a(\bar{z})~,\nn
&& \hspace{12cm}~~~ (\ep_i=0,1)~.
\label{bc fermion}
\end{eqnarray}
This condition is most easily understood by recalling 
the correct boundary conditions
in the thermal field theory of point particles 
(identified as the $w=0$ sector)
and further taking account of the consistency with modular invariance.

In this way
the desired partition function is calculated in the following form; 
\begin{eqnarray}
Z_{\msc{torus}}(\beta)&=& \int_{\cF}\frac{d^2\tau}{\tau_2^2}\,\nu 
\sum_{\ep_i=0,1} \,\sum_{\stackrel{w\in 2\bsz+\ep_1}{n\in 2\bsz+\ep_2}}
\,\sum_{r,s}\,e^{-S_{\msc{inst}}(w,n,r,s)}\,
Z^{\msc{tr}}_{\ep_1,\ep_2}(\tau,\bar{\tau}) ~,
\label{part flat 1}\\
Z^{\msc{tr}}_{\ep_1,\ep_2}(\tau,\bar{\tau})&=&
\frac{V_8}{(4\pi^2\al'\tau_2)^4} \frac{1}{|\eta(\tau)|^{16}}\cdot 
e^{-2\pi \tau_2\ep_1} 
\left|\frac{\th_1(\tau,\frac{\ep_1}{2}\tau+\frac{\ep_2}{2})}
{\eta(\tau)}\right|^8
~.
\end{eqnarray}
The subscripts $\ep_1$, $\ep_2$ indicate the twisted boundary
conditions of GS fermions \eqn{bc fermion}.
The calculation of transverse partition function 
$Z^{\msc{tr}}_{\ep_1,\ep_2}(\tau,\bar{\tau})$ 
is quite familiar, and we denoted the transverse volume factor as $V_8$. 
The integration region 
$\cF$ is the familiar fundamental domain 
\begin{equation}
\cF \df \left\{\tau \in \bc~;~ \tau_2>0,~|\tau|>1,~ 
|\tau_1|\leq \frac{1}{2}\right\}~.
\end{equation}

In the summation of winding numbers the $w=n=0$ sectors lead 
to a divergent term that should be interpreted as the vacuum
energy and does not depend on the parameters $\beta$, $R_-$.
We shall thus subtract them and assume $(w,n)\neq (0,0)$
to define the thermal partition function.
Substituting the expression \eqn{inst action 1},
we readily carry out the summation  over $r$, $s$, providing 
a periodic delta function.
The result is written as 
\begin{eqnarray}
Z_{\msc{torus}}(\beta)&=& \int_{\cF}\frac{d^2\tau}{\tau_2^2}\,
\nu \sum_{\ep_i=0,1} \,\sum_{\stackrel{w\in 2\bsz+\ep_1}{n\in 2\bsz+\ep_2}}
\,\sum_{p,q}\, 
e^{-\frac{\beta^2|w\tau-n|^2}{4\pi\al'\tau_2}}  \nn
&& ~~~ \times
\tau_2 \delta^{(2)}\left((w\nu+ip)\tau-(n\nu+iq)\right)
\,
Z^{\msc{tr}}_{\ep_1,\ep_2}(\tau,\bar{\tau}) ~,
\label{part flat 2}
\end{eqnarray}
which coincides with that given in \cite{Semenoff} calculated 
based on the RNS formalism.
The appearance of periodic delta function 
discretizes the moduli space of torus, which is a characteristic feature
of the DLCQ string theory.  The partition function \eqn{part flat 2} is 
manifestly modular invariant, and hence it is a  correct form of 
string amplitude on torus.

~

Let us turn to the pp-wave case.
At first glance we are likely to face an apparent difficulty,
since the light-cone gauge $X^+ \propto \tau$ is not compatible 
with the boundary conditions \eqn{instanton} in general. So,
one might be afraid that the light-cone gauge quantization would fail
in the thermal model.
However, we can instead take a natural gauge condition
$X^+=X^+_{w,n,r,s}$, where $X^+_{w,n,r,s}$ 
denotes the instanton solution for the each topological sector \eqn{instanton}.
We can further consider the rotation of world-sheet coordinates;
$z'= e^{i\th_{w,n}}z$, $\bar{z}'=e^{-i\th_{w,n}}\bar{z}$ with
\begin{equation}
\cos \th_{w,n}=\frac{\tau_1w-n}{|\tau w-n|}~,~~~
\sin \th_{w,n}=-\frac{\tau_2 w}{|\tau w-n|}~,
\label{rotation}
\end{equation}
such that 
\begin{eqnarray}
(\partial_{z'}+\partial_{\bar{z}'}) X^+_{w,n,r,s}\equiv 0~,~~~
(\partial_{z'}-\partial_{\bar{z}'}) X^+_{w,n,r,s}\equiv
- \frac{\sqrt{2}\beta}{4\pi\tau_2}|w\tau-n|~.
\label{instanton gauge}
\end{eqnarray}
Working with the new coordinates $z'$, $\bar{z}'$,
we still obtain the quadratic action \eqn{lc action} with 
the mass parameter 
\begin{eqnarray}
m=\frac{\sqrt{2}\mu \beta}{4\pi \tau_2}|w\tau-n|\equiv
\hm \frac{\nu}{\tau_2}|w\tau-n|~,
\end{eqnarray}
where we set $\hm = \mu \al'/R_-$.
To be more precise, because the complex structure defined by $z'$, $\bar{z}'$
depends on the choice of $w$, $n$, which may make the quantization
problematic, we must go back to the original coordinates $z$, $\bar{z}$
after making the gauge fixing. 
Fortunately, nothing is changed by this rotation
because of the manifest rotational symmetry of the action \eqn{lc action}.
We will later face a more non-trivial situation, in which we need a
care for such a rotation of complex coordinates, in the analysis of 
the 6-dimensional pp-wave. 

In this way we have found that the transverse partition function
for the each topological sector 
is evaluated by using the quadratic action \eqn{lc action}
as in the previous subsection, but with the non-trivial
mass parameter $\dsp m=\hm \frac{\nu}{\tau_2}|w\tau-n|$  
{\em depending on the thermal winding numbers $w$, $n$\/.} 
The desired partition function is thus calculated as;
\begin{eqnarray}
Z_{\msc{torus}}(\beta)&=& \int_{\cF}\frac{d^2\tau}{\tau_2^2}\,\nu 
\sum_{\ep_i=0,1} \,\sum_{\stackrel{w\in 2\bsz+\ep_1}{n\in 2\bsz+\ep_2}}
\,\sum_{r,s}\,e^{-S_{\msc{inst}}(w,n,r,s)}\,
Z^{\msc{tr}}_{\ep_1,\ep_2}\left(\tau,\bar{\tau};
\hm \frac{\nu}{\tau_2}|w\tau-n|\right) ~.
\label{part pp 0}
\end{eqnarray}
Since the transverse partition functions $Z^{\msc{tr}}_{\ep_1,\ep_2}$
do not depend on the windings $r$, $s$, we can likewise make the summation 
over them, yielding the same periodic delta function.
Recalling the results \eqn{Ztr 2}, \eqn{Ztr 3}, we finally obtain
\begin{eqnarray}
Z_{\msc{torus}}(\beta)&= & \int_{\cF}\frac{d^2\tau}{\tau_2^2}\,
\nu \sum_{\ep_i=0,1}\sum_{\stackrel{w\in 2\bsz+\ep_1}{n\in2\bsz+\ep_2}}
\sum_{p,q}\,
e^{-\frac{\beta^2|w\tau-n|^2}{4\pi\al'\tau_2}}
\tau_2 \delta^{(2)}\left((w\nu+ip)\tau-(n\nu+iq)\right) \nn
&& \hspace{1in} \times 
\frac{\Theta_{(\frac{1}{2}\ep_1,\frac{1}{2}\ep_2)}(\tau,\bar{\tau}
;\hat{\mu}|w\nu+ip|)^4}
{\Theta_{(0,0)}(\tau,\bar{\tau}
;\hat{\mu}|w\nu+ip|)^4}~.
\label{part pp}
\end{eqnarray}
Here we made use of the replacement of the mass parameter
$\dsp \hm \frac{\nu}{\tau_2}|w\tau-n|$ with the simpler one 
$\hm |w\nu + ip|$ due to the constraints  
\begin{eqnarray}
&& w\nu \tau_1-p\tau_2=n\nu~, \nn
&& w\nu \tau_2+p\tau_1=q~,
\label{delta constraints}
\end{eqnarray} 
imposed by the delta function factor.
This partition function \eqn{part pp} has 
the manifestly modular invariant form.
In fact, we can directly check it
by means of the evaluation 
\begin{eqnarray}
&&Z_{\ep_1,\ep_2}^{\msc{tr}}(-1/\tau,-1/\bar{\tau};\hat{\mu}|w\nu+ip|) 
\times \frac{\tau_2}{|\tau|^2}
\delta^{(2)}\left((w\nu+ip)(-1/\tau)-(n\nu+iq)\right) \nn
&& ~~~
=Z_{\ep_2,\ep_1}^{\msc{tr}}(\tau,\bar{\tau};\hat{\mu}|w\nu+ip||\tau|^{-1}) 
\times \tau_2
\delta^{(2)}\left((n\nu+iq)\tau+(w\nu+ip)\right) \nn
&& ~~~
= Z_{\ep_2,\ep_1}^{\msc{tr}}(\tau,\bar{\tau};\hat{\mu}|n\nu+iq|)
\times \tau_2
\delta^{(2)}\left((n\nu+iq)\tau+(w\nu+ip)\right)~.
\end{eqnarray}

We next present  the calculation by operator formalism, 
which will justify the correctness of our partition function \eqn{part pp}.

~

\subsection{Free Energy of Space-time Theory : Operator Calculation}
\indent

In general, the free energy (or the grand potential with the vanishing 
chemical potential) in the thermal ensemble of free string theory
is computed as
\begin{eqnarray}
F&=&\frac{1}{\beta}
\tr \left\lb (-1)^{\msc{\bf F}}
\ln \left(1-(-1)^{\msc{\bf F}}e^{-\beta p^0}\right)\right\rb \nn
&\equiv& -\sum_{n=1}^{\infty}\frac{1}{\beta n}
\tr \left\lb (-1)^{(n+1)\msc{\bf F}}e^{-\beta n p^0}\right\rb ~,
\label{free energy}
\end{eqnarray}
where $\mbox{\bf F}$ denotes the space-time fermion number (mod 2)
and $\dsp p^0 \equiv \frac{1}{\sqrt{2}}(p^+ - p^-)$ 
is the space-time energy operator. The trace should be 
taken over the single particle physical Hilbert space on which 
the on-shell condition and the level matching condition are imposed.
The free energy $F$ should be identified with the one-loop 
partition function of the first quantized thermal string
$Z_{\msc{torus}}(\beta)$ we studied above,  by the next simple relation
\begin{equation}
Z_{\msc{torus}} = -\beta F~.
\label{Z and F}
\end{equation}
The main aim in this subsection is to confirm this relation for 
the partition function \eqn{part pp}.

We start with the simple identity derived from the on-shell condition;
\begin{eqnarray}
p^0 =  \frac{1}{\sqrt{2}}\left(p^+-p^- \right)
= \frac{1}{\sqrt{2}}\left(\frac{p}{R_-}+\frac{R_-}{\al'p}H \right) ~,
\end{eqnarray}
where $H\equiv \al'p^+ H_{\msc{l.c.}}$ denotes the world-sheet 
Hamiltonian as before.
To impose the level matching condition \eqn{level matching} 
$\dsp P(\equiv \sum_n nN_n)=pk$ 
(${}^{\forall} k \in \bz$), it is convenient to insert
the following projection operator into the trace; 
\begin{equation}
\frac{1}{p}\sum_{q\in\bsz_p}\,e^{2\pi i \frac{q}{p}P}~.
\end{equation}
We so obtain the following expression from \eqn{free energy};
\begin{eqnarray}
F(\beta) = -\sum_{n=1}^{\infty}\sum_{p=1}^{\infty}\sum_{q\in \bsz_p}\,
\frac{1}{\beta n p} e^{-\frac{\beta n p}{\sqrt{2}R_-}} 
\tr \left\lb (-1)^{(n+1)\bsF} 
e^{-\beta n \frac{R}{\sqrt{2}p\al'}H +2\pi i \frac{q}{p}P}\right\rb~.
\label{free energy 1}
\end{eqnarray}
It is also convenient to introduce the ``modulus parameter'' 
$\dsp \tau \equiv \frac{q+in\nu}{p}$, where we set $\dsp \nu \equiv 
\frac{\sqrt{2}\beta R_-}{4\pi \al'}$ as before. 
\eqn{free energy 1} is rewritten as 
\begin{eqnarray}
F(\beta) = -\sum_{n,p,q}\,
\frac{1}{\beta n p} e^{-\frac{\beta^2 n^2}{4\pi \al' \tau_2}} 
\tr \left\lb (-1)^{(n+1)\bsF} 
e^{-2\pi \tau_2 H +2\pi i \tau_1 P} \right\rb~,
\label{free energy 2}
\end{eqnarray}
where the integers $n,p(>0),q$ run over the range such that 
$\tau \in \cS$ with the definition
\begin{equation}
\cS \df \left\{\tau \in \bc~;~ \tau_2>0,~ 
|\tau_1|\leq \frac{1}{2}\right\}~.
\end{equation}
The trace in this expression \eqn{free energy 2} 
is already calculated in \eqn{Ztr 2}, \eqn{Ztr 3}.
We thus finally obtain 
\begin{eqnarray}
F(\beta) = -\sum_{p,q}\left\lb \sum_{n:\msc{even}}\,
\frac{1}{\beta n p} e^{-\frac{\beta^2 n^2}{4\pi \al' \tau_2}} 
+ \sum_{n:\msc{odd}}\,
\frac{1}{\beta n p} e^{-\frac{\beta^2 n^2}{4\pi \al' \tau_2}}
\frac{\Theta_{(0,\frac{1}{2})}(\tau,\bar{\tau};\hat{\mu}p)^4}
{\Theta_{(0,0)}(\tau,\bar{\tau};\hat{\mu}p)^4}
\right\rb ~.
\label{free energy 3}
\end{eqnarray}

Let us now compare this result with the modular invariant 
partition function \eqn{part pp}. For this purpose 
it is easiest to make use of the technique invented
in \cite{Polchinski}. 
We first note that \eqn{part pp} has the form such as 
\begin{equation}
Z_{\msc{torus}}(\beta)
=\sum_{w,n}\int_{\cF}\frac{d^2\tau}{\tau_2^2} f_{(w,n)}(\tau,\bar{\tau})~,
\end{equation}
where $w$, $n$ denote the thermal winding numbers defined in \eqn{instanton},
and behave as the doublet of $PSL(2;\bz)$ under the modular transformations.
Moreover, we can always find out a modular transformation
setting $w=0$ for arbitrary $(w,n) \neq (0,0)$. 
Therefore, because of the modular invariance, we can simply set 
$w=0$ in the integrand of \eqn{part pp}, but must replace
the fundamental domain $\cF$ with the larger domain $\cS$ defined above.
In summary,  the partition function \eqn{part pp} can be rewritten in
the following simpler form,  although we loose the manifest 
modular invariance;
\begin{eqnarray}
Z_{\msc{torus}}(\beta)&= & \int_{\cS}\frac{d^2\tau}{\tau_2^2}\,
\nu \sum_{\ep=0,1} \sum_{n\in 2\bsz + \ep}\sum_{p,q}\,
e^{-\frac{\beta^2 n^2}{4\pi\al'\tau_2}}
\tau_2 \delta^{(2)}\left(ip\tau-(n\nu+iq)\right) \,
\frac{\Theta_{(0,\frac{1}{2}\ep)}(\tau,\bar{\tau}
;\hat{\mu}|p|)^4}
{\Theta_{(0,0)}(\tau,\bar{\tau}
;\hat{\mu}|p|)^4}~\nn
&=& \sum_{p,q}\left\lb \sum_{n\,:\,\msc{even}}\,
\frac{1}{n p} e^{-\frac{\beta^2 n^2}{4\pi \al' \tau_2}} 
+ \sum_{n\,:\,\msc{odd}}\,
\frac{1}{n p} e^{-\frac{\beta^2 n^2}{4\pi \al' \tau_2}}
\frac{\Theta_{(0,\frac{1}{2})}(\tau,\bar{\tau};\hat{\mu}p)^4}
{\Theta_{(0,0)}(\tau,\bar{\tau};\hat{\mu}p)^4}
\right\rb ~.
\label{part pp 2}
\end{eqnarray}
In the last line we set $\dsp \tau = \frac{q+in\nu}{p}$, and 
the summation with respect to $n,p,q$ should be taken 
over the range such that $\tau \in \cS$.
The relation \eqn{Z and F} is obviously confirmed.
Therefore, the validity of the partition function \eqn{part pp} 
has been confirmed.

It is worth pointing out that the first term (the summation over even
$n$) in \eqn{part pp 2} is ``topological'' one  originating from  
the Witten index counting the BPS states. This term is absent in 
the case of flat background. In fact, consider the flat limit 
$\mu \,\rightarrow\,0$. There emerges a divergent volume factor $V_8$
from the second term and thus the first term becomes negligible.
The partition function per unit volume has the next limit  
\begin{eqnarray}
\lim_{\mu\rightarrow 0}\frac{Z_{\msc{torus}}(\beta)}{V_8}
= \sum_{p,q}\sum_{n\,:\,\msc{odd}}\, 
\frac{1}{n p}e^{-\frac{\beta^2 n^2}{4\pi \al' \tau_2}}\,
\frac{1}{(4\pi^2 \al' \tau_2)^4}\frac{1}{|\eta(\tau)|^{16}}\cdot
\left|\frac{\th_2(\tau)}{\eta(\tau)}\right|^8~,
\end{eqnarray}
which is identical to the thermal partition function
(per unit volume) 
in the DLCQ flat background calculated in \cite{Semenoff}.



We also comment on the decompactification limit 
$R_-\,\rightarrow\,\infty$. To consider it, it is the easiest to 
start from  \eqn{part pp 0}. Under this limit the DLCQ windings 
$r$, $s$ decouple, and we obtain
\begin{eqnarray}
&& 
\lim_{R_-\rightarrow\infty} \, \frac{1}{\sqrt{2}\pi R_-}Z_{\msc{torus}}(\beta)
= \int_{\cF}\frac{d^2\tau}{\tau_2^2}\,
\frac{\beta}{4\pi^2\al'}\, \sum_{\ep_i=0,1}\,
\sum_{\stackrel{w\in 2\bsz+\ep_1}{n\in 2\bsz+\ep_2}}\,
e^{-\frac{\beta^2|w\tau-n|^2}{4\pi\al'\tau_2}}\,    \nn
&& \hspace{1in} \times Z^{\msc{tr}}_{\ep_1,\ep_2}\left(\tau,\bar{\tau};
\frac{\sqrt{2}\mu\beta}{4\pi\tau_2}|w\tau-n|\right)~.
\label{part pp dec 1}
\end{eqnarray}
Expressing $Z^{\msc{tr}}_{\ep_1,\ep_2}$ 
by the appropriate massive theta functions 
as before, 
we achieve the modular invariant form of partition function.
We can also rewrite it by setting $w=0$ and replacing the integration 
region $\cF$ with  $\cS$ based on the same argument. 
Transforming the integration variable as 
$\dsp p^+=\frac{\sqrt{2}\beta n}{4\pi \al'\tau_2}$, we obtain
\begin{eqnarray}
\lim_{R_-\rightarrow\infty}  \, \frac{1}{\sqrt{2}\pi R_-}
Z_{\msc{torus}}(\beta)
= \sum_{\ep=0,1}\, \sum_{n\in 2\bsz+\ep,\,n>0}\, \frac{1}{n}\,
\int_0^{\infty}\frac{dp^+}{\sqrt{2}\pi} 
\int_{-\frac{1}{2}}^{\frac{1}{2}}d\tau_1\,
e^{-\frac{\beta n p^+}{\sqrt{2}}}\,
Z^{\msc{tr}}_{0,\ep}\left(\tau,\bar{\tau};
\mu \al' p^+\right),
\label{part pp dec 2}
\end{eqnarray}
where we wrote $\dsp \tau=\tau_1+i\frac{\sqrt{2}\beta n}{4\pi \al'p^+}$.
The last expression \eqn{part pp dec 2} is readily compared with 
the free energy evaluated by the operator formalism and also 
corresponds to that presented in \cite{PV,GSS}. 

~

 
\subsection{``Virtual String'' and Hagedorn Temperature}
\indent

Let us return to the modular invariant partition function \eqn{part pp}.
As is clear from the above analysis, the sectors of $w=0$, 
$p\in \bz_{>0}$ correspond to the states in the physical Hilbert
space of the non-thermal string theory 
(with the Lorentzian signature)
and the integer $p$ is precisely identified as the light-cone 
momentum by the relation $p^+=p/R_-$. This is a natural 
correspondence since $p$, $q$ are actually the ``momenta''
dual to the DLCQ winding numbers $r$, $s$. The strings 
in these physical sectors have the standard 
mass parameters $m= \al' \mu p^+\equiv \hm p$ 
and satisfy the on-shell condition $-\al'p^+p^-=H$, where $H$
is the world-sheet Hamiltonian $H\equiv \sum_{n\in \bsz}\om_n N_n$,
as well as the level matching condition \eqn{level matching}.

However, we still have the sectors with the non-vanishing
(spatial) thermal windings $w\neq 0$. They do not correspond
to any physical states in the non-thermal theory. In this sense
we shall call them  the ``virtual strings'' throughout this
paper\footnote
        {The partition function \eqn{part pp} 
   including the contributions $w\neq 0$ has been derived by 
   the path-integral calculation based on the complex world-sheet 
   action, as in the flat DLCQ case \cite{Semenoff}. 
   So, one should regard them as ``virtual'' in the doubly
   meaning. Such virtual winding modes are surely useful to gain 
   the manifest modular invariance and a concise understanding of 
   the thermal instability as discussed here.  The physical meaning of 
   the virtual strings resides in this fact. }. 
The world-sheet theories for  the virtual strings are likewise described 
by the world-sheet Hamiltonian $H$, but 
{\em  with the modified mass parameter $m= \hm|w\nu+ip|$.}
The transverse partition function is calculated in the same way and
composes the building blocks of the modular invariant amplitude
\eqn{part pp}.

In summary, the manifest modular invariance in \eqn{part pp} 
requires  the contributions from 
the various virtual string sectors $w\neq 0$, 
while the alternative expression \eqn{part pp 2} 
only contains the physical string states, 
although the modular invariance is hidden. 
This is a general feature of the thermal string theory.

The virtual string states could be tachyonic in spite of 
the unbroken space-time SUSY, owing to the lack of mass-shell
condition in the usual sense. 
This fact leads to a simple interpretation of Hagedorn 
temperature \cite{Hagedorn} as explained in \cite{Sath,AW}. 
In fact, we can make use of the analogous
argument based on the modular invariance of \eqn{part pp}
in the previous subsection, but employ the different
gauge choice $n=0$ in \eqn{part pp} rather than $w=0$. 
In that case the amplitude is UV finite ($\tau_2 \sim 0$), 
but could have a IR divergence ($\tau_2 \sim + \infty$)
due to the tachyonic mode. 
It is not difficult to see that the leading term is  
the contribution from the virtual string with $w=1$, $n=0$,
which has the mass parameter 
$\dsp m=\hm \nu\equiv \frac{\sqrt{2}\mu\beta}{4\pi}$.  
We thus find that 
\begin{eqnarray}
Z_{\msc{torus}}(\beta)~:~\mbox{finite}~\Longleftrightarrow~
\frac{\beta^2}{8\pi^2\al'}> 8\left(
\Delta(\frac{\sqrt{2}\mu\beta}{4\pi};\frac{1}{2})
- \Delta(\frac{\sqrt{2}\mu\beta}{4\pi};0)\right) ~.
\label{H ineqality 1}
\end{eqnarray}
It is easy to see that the R.H.S of the inequality is always positive
and a monotonically decreasing function of $\beta$.
Thus we can rewrite it as
\begin{eqnarray}
Z_{\msc{torus}}(\beta)~:~\mbox{finite}~\Longleftrightarrow~
\beta> \beta_H~,
\label{H ineqality 2}
\end{eqnarray}
with
\begin{eqnarray}
\frac{\beta_H^2}{8\pi^2\al'}-8\left(
\Delta(\frac{\sqrt{2}\mu\beta_H}{4\pi};\frac{1}{2})
- \Delta(\frac{\sqrt{2}\mu\beta_H}{4\pi};0)\right) =0~.
\label{H temperature 1}
\end{eqnarray}
the critical temperature $T_H\equiv \beta_H^{-1}$ is no other than the
Hagedorn temperature at which the thermal instability occurs.
This has the correct limit under $\mu\,\rightarrow\,0$
\begin{eqnarray}
\lim_{\mu \rightarrow 0}T_H = \frac{1}{\sqrt{8\pi^2\al'}}~,
\label{H temperature flat}
\end{eqnarray}
consistent with the result given in \cite{AW}.

An alternative interpretation of such thermal instability is presented 
from the ``dual'' expression \eqn{part pp 2} that only includes the physical 
states. This is clearly IR finite, but could be UV divergent due to 
the rapid growth of massive excitations depending exponentially
on the oscillator level.

We further make a few comments.

Firstly, the Hagedorn temperature $T_H$ does not depend on the DLCQ radius
$R_-$ as in the flat background. 
In fact, the equation \eqn{H temperature 1} is 
equivalent with those given in the recent papers
\cite{PV,GSS} (with the suitable identification of parameters), 
in which the analysis  for the non-DLCQ thermal model is presented.

Secondly, because the R.H.S. of \eqn{H temperature 1} 
is a monotonically decreasing function of $\mu$, 
it is easy to see that $T_H$
is bounded from below by the value for the flat background ($\mu=0$);
\begin{equation}
 T_H \geq T_{H,\,\msc{flat}} \equiv \frac{1}{\sqrt{8\pi^2\al'}}~,
\end{equation}
and diverges under the large $\mu$ limit.
This means that the Hagedorn transition does not occur at any
finite temperature under this limit. 
The stringy nature is expected to be lost under the large $\mu$ limit
so that the picture of ``string bit'' becomes a good approximation
\cite{string bit,Verlinde}. Such behavior of Hagedorn temperature 
is likely to be consistent with this aspect. 

~

\section{Thermal Amplitudes of Open Strings in DLCQ PP-Waves}
\indent

As an extension of our previous analysis, let us 
consider the thermal ensemble of open strings in the DLCQ pp-waves
with supersymmetric D-branes (half BPS D-branes, strictly speaking).
We shall only focus on the D-branes in  the maximally supersymmetric 
10-dimensional pp-wave \cite{BP,DP,BGG}, although the generalization to
more general backgrounds 
(say, the 6-dimensional pp-waves analyzed in the next section) 
is straightforward (see the papers \cite{pp brane}).

Turning our attention to the open-closed string duality 
in cylinder amplitudes (or ``Cardy condition''),  
we seem to face a difficulty originating
from the light-cone gauge quantization. 
For example, pick up the time-like D-branes that impose the Neumann
boundary condition along the light-cone directions $X^+$, $X^-$.
In this case the open string picture is compatible with the light-cone
gauge $X^+=2 \al' p^+\tau$ \footnote
     {It is natural to define the light-cone gauge for open string 
      as $X^+=2 \al' p^+\tau$ rather than $X^+=\al' p^+\tau$.
      The parameter $p^+$ in this expression is really identified as the 
      momentum canonically conjugate to the zero-mode variable $x^-$
      in the case of open string, which should be quantized as 
      $p^+=p/R_-$ ($p\in \bz_{>0}$) in DLCQ. The difference 
      of factor is originating from  the simple fact 
      that the spatial direction of open 
      string world-sheet is parametrized as $0\leq \sigma\leq \pi$,
      while that of the closed string is done as $0\leq \sigma < 2\pi$
      in our convention.}. 
However, since the boundary conditions
for the closed string channel imply   $\partial_{\tau}X^+=0$, 
we cannot choose the usual light-cone gauge. 
In the case of Euclidean D-brane (or, the D-brane instantons)
the situation is opposite. Namely, the closed string description is 
compatible with the light-cone gauge, while 
we cannot take it in the open string channel.
This naive observation may look puzzling  and leads to 
an apparent discrepancy between the classifications of supersymmetric
D-branes given in \cite{DP} and \cite{BP}, where the former is based on 
the open string picture and the latter is the boundary 
state approach (closed string picture).

A clear resolution to this puzzle for the thermal model 
is given by considering the virtual strings we discussed above.
Although the virtual strings  are not compatible 
with the light-cone gauge condition, 
we can consistently define the world-sheet Hamiltonian in the quadratic form 
by taking the ``instanton gauge'' as before. 

We shall separately discuss the cases of the time-like D-branes and
the Euclidean D-branes.

~


\subsection{Thermal Cylinder Amplitude for Time-like D-branes} 
\indent

We first consider the time-like Dp-branes, with which 
the light-cone coordinates $X^+$, $X^-$ should satisfy 
the Neumann boundary conditions. 
As is shown in \cite{DP}, the supersymmetry condition 
(for the half BPS brane)
implies that only the cases $p=3,5,7$ are allowed and 
the branes must be stacked at the origin of transverse plane $X^I=0$
(if not assuming the extra flux).
We only focus on the simplest configurations such that the open strings have
the both ends attached to the same Dp-brane, 
which are manifestly supersymmetric.

The basic aspects are summarized as follows:
\begin{description}
 \item[1. open string channel]

~

We have the physical strings compatible with the light-cone gauge
condition $\dsp X^+= 2 \al' p^+\tau \equiv 2 \al'  \frac{p}{R_-}\tau$ 
and satisfying the mass-shell condition. The world-sheet Hamiltonian
$H^{\msc{(o)}}$ includes the standard mass parameter $m=2\mu \al' p^+ 
\equiv 2 \hm p$.

\item[2. closed string channel]

~

The boundary states only contain the virtual string states
not compatible with the usual light-cone gauge condition and  
not satisfying the mass-shell condition.
The world-sheet Hamiltonian $H^{\msc{(c)}}$
includes the mass parameter $m=\hm w \nu$ as we will see below. 

\end{description}


We begin with the calculation in the open string channel.
As in the calculation of closed string partition function,
we employ the path-integral approach (especially for 
the longitudinal sector).

As a preparation we introduce
the massive theta functions 
for the open string amplitudes;
\begin{eqnarray}
\th_{(a,b)}(t;m)& \df& \sqrt{\Theta_{(a,b)}(it,-it;m)} \nn
& \equiv & e^{2\pi t \Delta(m;a)}\,\prod_{n\in \bsz}
\left| 1- e^{-2\pi t \sqrt{m^2+(n+a)^2} +2\pi i b }\right|
\label{open theta}
\end{eqnarray} 
where $t>0$ is the open string modulus\footnote
      {The ``modified f-functions'' defined in \cite{BGG} correspond to 
    $\th_{(0,0)}(t;m)^{1/2}$, $\th_{(0,1/2)}(t;m)^{1/2}$, 
    $\th_{(1/2,1/2)}(t;m)^{1/2}$, and $\th_{(1/2,0)}(t;m)^{1/2}$.}.
They have the following modular property
\begin{eqnarray}
\th_{(a,b)}(1/t;mt)=\th_{(b,-a)}(t;m)~,
\label{modular th}
\end{eqnarray}
and the mass-less limits
\begin{eqnarray}
\lim_{m\rightarrow 0}\th_{(a,b)}(t;m) = e^{-\pi t a^2}\,
\left|\frac{\th_1(it,iat+b)}{\eta(it)}\right|~.
\end{eqnarray} 

We first consider the bosonic amplitudes in the transverse sector.
Recall that the NN open string has the zero-modes, while the DD open string 
does not. The zero-point energy (per one {\em complex} boson)
is thus evaluated as 
\begin{eqnarray}
m+\sum_{n=1}^{\infty}\sqrt{m^2+n^2}-\int_0^{\infty}dk\,\sqrt{m^2+k^2}
 =  \frac{m}{2}+\Delta(m;0)~,
\label{zero point NN}
\end{eqnarray}
for the NN strings,  and 
\begin{eqnarray}
\sum_{n=1}^{\infty}\sqrt{m^2+n^2}-\int_0^{\infty}dk\, \sqrt{m^2+k^2}
= -\frac{m}{2}+\Delta(m;0)~,
\label{zero point DD}
\end{eqnarray}
for the DD strings. 
Since we have the $\mbox{p}-1$ components of 
NN strings and $9-\mbox{p}$ components of DD strings 
along the transverse plane, the bosonic amplitude 
is evaluated as follows \cite{BGG};
\begin{eqnarray}
 \tr_{\cH_b}\left\lb e^{-2\pi t H^{\msc{(o)}}} \right\rb
&=& \frac{q^{\frac{m}{2}\left(\frac{\msp-1}{2}-\frac{9-\msp}{2}\right)}}
{(1-q^m)^{\msp-1}} \times \frac{1}{\th_{(0,0)}(t;m)^4/(1-q^m)^4} \nn
&=& \left(2 \sinh (\pi t m)\right)^{5-\msp} \cdot 
\frac{1}{\th_{(0,0)}(t;m)^4}~,
\label{bosonic open amplitude 1} 
\end{eqnarray} 
where we wrote  $q \equiv e^{-2\pi t}$.

For the fermionic amplitudes, a careful analysis is needed for
the zero-modes. This is presented  in \cite{BGG} and the results are quite 
simple;
\begin{eqnarray}
\tr_{\cH_f}\left\lb (-1)^{\bsF}\,e^{-2\pi t H^{\msc{(o)}}}\right\rb 
&=& \th_{(0,0)}(t;m)^4~,\nn
\tr_{\cH_f}\left\lb e^{-2\pi t H^{\msc{(o)}}}\right\rb 
&=& \th_{(0,1/2)}(t;m)^4~.
\label{fermionic open amplitude 1}
\end{eqnarray}

Let us next consider the longitudinal sector.
We have various topological sectors characterized by the windings 
$n$, $s$;
\begin{eqnarray}
&&  X^+(\tau_E+2\pi t,\sigma) = X^+(\tau_E,\sigma)
+i\frac{\beta n}{\sqrt{2}}~, \nn
&&  X^-(\tau_E+2\pi t, \sigma) = X^-(\tau_E,\sigma)-i\frac{\beta n}{\sqrt{2}}
+ 2\pi R_-s~, \nn
&& \partial_{\sigma} X^+|_{\sigma=0,\pi} 
= \partial_{\sigma} X^-|_{\sigma=0,\pi}=0~.
\label{instanton 2}
\end{eqnarray}
As in the previous analysis, we shall make 
the ``instanton gauge''  $X^+=X^+_{n,s}$,
where the $X^+_{n,s}$ is the instanton characterized  by the above boundary 
condition \eqn{instanton 2}. It makes the world-sheet action
quadratic. The mass parameter is evaluated as 
$\dsp m= \mu \left|\frac{\partial X^+_{n,s}}{\partial \tau_E}\right| 
\equiv \hm n\nu/t$.  The instanton action is also calculated as 
\begin{eqnarray}
S_{\msc{inst}}(n,s)
= \frac{\beta^2 n^2}{8\pi \al' t}+ 2\pi i \frac{\nu}{2t} ns~.
\end{eqnarray}
So, the longitudinal amplitude becomes
\begin{eqnarray}
&& \frac{V_{\msc{l.c.}}}{8\pi^2 \al' t} \times \sum_{n,s}
e^{-S_{\msc{inst}}(n,s)}
\equiv \frac{\nu}{2t}\, \sum_{n,s}e^{-S_{\msc{inst}}(n,s)}~.
\end{eqnarray}
where the prefactor is again equal to the Gaussian integral
of zero-mode momenta in the Hamiltonian formalism.

Gathering all the things, we achieve the following 
thermal cylinder amplitude
\begin{eqnarray}
 Z^{\msc{(o)}}_{\msc{cylinder}}(\beta;\mbox{Dp})
&=& \int_0^{\infty}\frac{dt}{t}\, \frac{\nu}{2}\,
\sum_{\ep=0,1}\, \sum_s\,\sum_{n\in 2\bsz+\ep}\,
e^{-S_{\msc{inst}}(n,s)}\, Z^{\msc{tr}}_{\ep}(t;\hm n \nu/t) \nn
&=& 
\int_0^{\infty}\frac{dt}{t}\,\frac{\nu}{2}\,
\sum_{\ep=0,1}\,\sum_p\,\sum_{n\in 2\bsz+\ep}\,
 e^{-\frac{\beta^2 n^2}{8\pi \al' t}}\, \delta (pt-n\nu/2) \nn
&& \hspace{1in} \times
\left(2 \sinh(2\pi t \hm p)\right)^{5-\msp}
\cdot \frac{\th_{(0,\frac{\ep}{2})}(t;2\hm p)^4}{\th_{(0,0)}(t;2\hm p)^4}~,
\label{part open 1} 
\end{eqnarray}
where $\ep$ indicates the thermal boundary condition of 
GS fermions. Again the DLCQ winding $s$ is dualized 
into the light-cone momentum $p$,  
yielding the periodic delta function. 
Notice that the correct mass parameter 
$m=2\mu \al'p^+ \equiv 2\hm p$ has been  successfully reproduced.
Performing the modulus integral explicitly, we also obtain
\begin{eqnarray}
&& Z^{\msc{(o)}}_{\msc{cylinder}}(\beta;\mbox{Dp})
=\sum_{p=1}^{\infty}\,\left\lb 
\sum_{n\,:\,\msc{even}, n>0}\,\frac{1}{n}e^{-\frac{\beta^2n^2}{8\pi \al' t}}
 \cdot \left(2 \sinh (2 \pi t \hm p)\right)^{5-\msp} \right. \nn 
&& \left. \hspace{1in} + \sum_{n\,:\,\msc{odd}, n>0}\,
\frac{1}{n}e^{-\frac{\beta^2n^2}{8\pi \al' t}}\cdot 
\left(2 \sinh (2 \pi t \hm p)\right)^{5-\msp}\cdot 
\frac{\th_{(0,\frac{1}{2})}(t;2\hm p)^4}{\th_{(0,0)}(t;2\hm p)^4}
\right\rb~,
\label{part open 2}
\end{eqnarray}
where we set $\dsp t= \frac{n\nu}{2p}$.
The last expression \eqn{part open 2} should coincide
with the open string free energy defined in the same way
as \eqn{free energy} (up to the factor $-\beta$).  
In fact, the on-shell condition 
for the open string is expressed as $H^{(\msc{o})}=-2\al' p^+p^-$,
and thus we have
\begin{eqnarray}
p^0=\frac{1}{\sqrt{2}} \frac{p}{R_-}+ \frac{R_-}{2\sqrt{2}\al'p}H^{\msc{(o)}}
~.
\label{p0 open}
\end{eqnarray}
Thanks to this relation, one can immediately confirm their coincidence. 
In particular  \eqn{part open 2} only includes the contributions from 
the physical open string states compatible with the light-cone 
gauge $\dsp X^+=2 \al' \frac{p}{R_-}\tau$.


~

Let us next analyze the closed string channel.
The wanted boundary state is composed only by the virtual string states 
that are not compatible with the light-cone gauge and 
do not satisfy the mass-shell condition\footnote
       {We now would like to emphasize that this is {\em not\/} a
        peculiarity of the pp-wave background. In fact, as the simplest
        example, let us recall the case in which the Neumann boundary
        conditions are imposed  
        along all the directions on  the non-compact flat space-time. 
        In this case, since the boundary state can only include  
        the zero momentum states, 
        the closed string states appearing in it  
        never satisfy the mass-shell condition (except for the massless 
         states with zero momenta).  Nevertheless,
        this boundary state is surely on-shell in the sense that the boundary 
        conformal symmetry is preserved; $L_n-\tilde{L}_{-n}=0$. 
        (${}^{\forall}n$)}.
It should have the following structure
\begin{eqnarray}
\ket{\mbox{Dp}} = \sum_{w,r}\, \cN_{w,r}\ket{w,r}\otimes 
\ket{\mbox{Dp};\hm w\nu}^{(\msc{tr})}~,
\label{b state 1}
\end{eqnarray}
where $\ket{w,r}$ are the zero-mode eigenstates for the longitudinal
directions  associated to the topological sectors
\begin{eqnarray}
&&  X^+(\tau_E, \sigma+2\pi) = X^+(\tau_E,\sigma)+i\frac{\beta w}{\sqrt{2}}
~, \nn
&&  X^-(\tau_E,\sigma+2\pi) = X^-(\tau_E,\sigma)-i\frac{\beta w}{\sqrt{2}}
+ 2\pi R_- r~, \nn
&& \partial_{\tau_E}X^+|_{\tau_E=0,\, \pi T}=
\partial_{\tau_E}X^-|_{\tau_E=0,\, \pi T}=0~.
\label{instanton 3}
\end{eqnarray}
$\ket{\mbox{Dp};\hm w\nu}^{(\msc{tr})}$ denotes the boundary states 
describing the transverse part of time-like Dp-brane,
which is similarly defined as that for the supersymmetric Euclidean
D(p-2)-brane presented in \cite{BP}, but with the mass parameter
$m=\hm w\nu$ instead of $m=\hm p$. 
The instanton configuration corresponding to
\eqn{instanton 3} indeed leads to the mass $m= \hm w\nu$
in the similar manner  as before.
Another important modification
is the thermal boundary condition of fermionic
fields. We must employ the integral modes of fermionic
oscillators for the even $w$, and the half-integral modes for the odd $w$.
$\cN_{w,r}$ are the numerical factors which should be determined 
by the requirements of open-closed duality.

The transverse part yields the standard overlap amplitude;
\begin{eqnarray}
&& {}^{(\msc{tr})}\!\bra{\mbox{Dp};\hm w \nu} (-1)^{\bsF}\,
e^{-\pi T H^{(\msc{c})}}
\ket{\mbox{Dp};\hm w\nu}^{(\msc{tr})} =
\left\{
\begin{array}{ll}
 \dsp \frac{\th_{(0,0)}(T;\hm w \nu)^4}{\th_{(0,0)}(T;\hm w \nu)^4}
   \equiv 1& ~~~ (w \in 2\bz)~, \\
 \dsp \frac{\th_{(\frac{1}{2},0)}(T;\hm w \nu)^4}
{\th_{(0,0)}(T;\hm w \nu)^4}
   & ~~~ (w \in 2\bz+1)~.
\end{array}
\right. 
\end{eqnarray}
The longitudinal part of closed string Hamiltonian $H_l^{(\msc{c})}$
is given as the zero-mode part of standard $L_0$ operator, which   
simply  gives 
\begin{eqnarray}
&& \bra{w,r}e^{-\pi T H_l^{(\msc{c})}} \ket{w',r'} = \delta_{w,w'}\delta_{r,r'}
e^{-S_{\msc{inst}}(w,r)}~, \nn
&& S_{\msc{inst}}(w,r) = \frac{\beta^2w^2T}{8\pi \al'} + \pi i T\nu wr~.
\label{instanton action 3}
\end{eqnarray}

In total, we obtain the cylinder amplitude in the closed string channel
\begin{eqnarray}
&& Z^{(\msc{c})}_{\msc{cylinder}}(\beta;\mbox{Dp}) \equiv 
\int_0^{\infty} dT\,
\bra{\mbox{Dp}} (-1)^{\bsF}\, e^{-\pi T H^{\msc{(c)}}} \ket{\mbox{Dp}} \nn
&& ~~~ = \int_0^{\infty}dT\,
\sum_{\ep=0,1} \sum_{w\in 2\bsz+\ep}\, \sum_r |\cN_{w,r}|^2\,
e^{-\frac{\beta^2w^2T}{8\pi \al'}+ \pi i T \nu wr} 
\cdot \frac{\th_{(\frac{\ep}{2},0)}(T;\hm w \nu)^4}
{\th_{(0,0)}(T;\hm w \nu)^4}
~.
\label{part closed 1}
\end{eqnarray}
We choose the normalization coefficients $\cN_{w,r}$ as
\begin{eqnarray}
\cN_{w,r} = \sqrt{\frac{\nu}{2}}\cdot
\left(2\sinh(\pi \hm w\nu)\right)^{\frac{5-\msp}{2}}~,
\label{N w r}
\end{eqnarray}
according to \cite{BGG}. The prefactor $\sqrt{\nu/2}$
is the correct one associated to the Neumann boundary
conditions along the longitudinal directions. 
Since $\cN_{w,r}$ does not depend on $r$,
we can  explicitly make the summation over $r$, resulting
\begin{eqnarray}
&& Z^{(\msc{c})}_{\msc{cylinder}}(\beta;\mbox{Dp}) 
= \int_0^{\infty}dT\,
\sum_{\ep =0,1 } \, \sum_{w\in 2\bsz + \ep}\, \sum_q 
e^{-\frac{\beta^2w^2T}{8\pi \al'}} \, \delta\left(
\frac{\nu}{2} w T-q\right)  \nn
&& \hspace{1in}  \times \frac{\nu}{2}\cdot
\left(2\sinh(\pi \hm w\nu)\right)^{5-\msp} \cdot
\frac{\th_{(\frac{\ep}{2},0)}(T;\hm w \nu)^4}
{\th_{(0,0)}(T;\hm w \nu)^4} ~.
\label{part closed 2}
\end{eqnarray}

The closed and open string moduli should be identified by the standard 
relation $t=1/T$. Comparing \eqn{part open 1} and \eqn{part closed 2},
and using the modular property of the massive theta function
\eqn{modular th}, one can find that
\begin{eqnarray}
&& Z^{(\msc{o})}_{\msc{cylinder}}(\beta;\mbox{Dp})
=Z^{(\msc{c})}_{\msc{cylinder}}(\beta;\mbox{Dp})~.
\end{eqnarray}
This equality  is no other than the wanted open-closed string duality.

We finally evaluate the Hagedorn temperature based on the cylinder 
amplitudes. As in the previous analysis of toroidal partition function,
it is easiest to evaluate the IR behavior of virtual strings,
namely, to study the behavior around $T\sim \infty$ in  
$Z^{(\msc{c})}_{\msc{cylinder}}(\beta;\mbox{Dp})$ for the present 
problem. Obviously the dominant term is $w=1$, and we find that
\begin{eqnarray}
Z^{(\msc{c})}_{\msc{cylinder}}(\beta;\mbox{Dp})~:~\mbox{finite}~
\Longleftrightarrow~ \beta > \beta_H~
\end{eqnarray}
with 
\begin{eqnarray}
\frac{\beta^2_H}{16\pi^2\al'}-4\left(\Delta
\left(\frac{\sqrt{2}\mu\beta_H}{4\pi};\frac{1}{2}\right)
-\Delta\left(\frac{\sqrt{2}\mu\beta_H}{4\pi};0\right)
\right)  =0~.
\label{H temperature 2}
\end{eqnarray}
It is equivalent with the equation \eqn{H temperature 1} for 
the closed string sector. 
Thus the open string sector has the equal Hagedorn temperature as in the 
case of flat background. Especially, it does not 
depend on the dimension of brane p.

~


\subsection{Thermal Ensemble of Closed String States 
Emitted/Absorbed by Euclidean D-branes}
\indent

We next consider the Euclidean Dp-branes, which impose the Dirichlet
condition for $X^+$, $X^-$. We shall express them as D'p to distinguish
from the time-like D-branes.
The compatibility with supersymmetry
implies the allowed values of p are $\mbp = 1,3,5$ \cite{BP},
and again they must be stacked at the origin in the transverse plane.

The roles played by the open and closed string channels are opposite
to the previous case;

\begin{description}
 \item[1. open string channel]

~

We have the virtual strings 
not compatible with the light-cone gauge.
The world-sheet Hamiltonian $H^{\msc{(o)}}$
includes the mass parameter $m=2 \hm w \nu$ as shown below.

\item[2. closed string channel]

~

The boundary states contain 
the physical string states  compatible with the light-cone gauge
condition $\dsp X^+=\al' p^+\tau \equiv \al' \frac{p}{R_-}\tau$. 
The world-sheet Hamiltonian
$H^{\msc{(c)}}$ includes the standard mass parameter 
$m=\mu \al' p^+ \equiv \hm p$.
\end{description}


We first discuss the closed string channel. 
Contrary to the previous analysis,
it might be ambiguous what type amplitude we should evaluate,
because  our purpose in this paper is to calculate the thermodynamical
trace {over  the physical states\/}. 
Generically, the overlap amplitudes of boundary states 
are not interpreted as  a trace over the closed string Hilbert space.
Nevertheless, it is quite natural to consider the next ``free energy''
\begin{eqnarray}
F(\beta;\mbox{D'p}) &\df& \frac{1}{\beta}\bra{\mbox{D'p}}
(-1)^{\bsF}\ln\left(1-(-1)^{\bsF}e^{-\beta p^0}\right) \ket{\mbox{D'p}}\nn
&\equiv & -\sum_{n=1}^{\infty}\, \frac{1}{\beta n}
\bra{\mbox{D'p}} (-1)^{(n+1)\bsF}\,e^{-\beta n p^0}\ket{\mbox{D'p}}~,
\label{free energy bs}
\end{eqnarray} 
where $\ket{\mbox{D'p}}$ denotes the boundary state describing the 
D'p brane localized at $X^+=X^-=0$ \footnote
    {In principle, we can consider the cylinder amplitude
     in which the ends of open string attached at the D'p branes 
     located at different points along the $X^+$, $X^-$ directions
     by including the suitable phase factors in $\cN'_p$.
      However, since our purpose is to calculate the {\em trace\/} 
      over the closed string states, we must consider the case of 
      open strings ended at the same brane.},
which should have the structure
\begin{eqnarray}
\ket{\mbox{D'p}} = \sum_{p,h}\, \cN'_{p}\, \ket{p^+=p/R_-}
\otimes \ket{p^-(p,h)}
\otimes \ket{\mbox{D'p},h;\hm p}^{(\msc{tr})}~.
\label{b state 2}
\end{eqnarray}
In this expression $\ket{\mbox{D'p},h;\hm p}^{(\msc{tr})}$
is defined by the decomposition of the transverse boundary state
of D'p brane $\ket{\mbox{D'p};\hm p}^{(\msc{tr})}$
with respect to the eigenvalue of $H^{(c)}$;
\begin{eqnarray}
&&\ket{\mbox{D'p};\hm p}^{(\msc{tr})}= \sum_h\,
\ket{\mbox{D'p},h;\hm p}^{(\msc{tr})} ~, \nn
&& H^{(c)}\ket{\mbox{D'p},h;\hm p}^{(\msc{tr})}=h
\ket{\mbox{D'p},h;\hm p}^{(\msc{tr})}~.
\end{eqnarray}
The longitudinal zero-mode parts $\ket{p^+=p/R_-}$, $\ket{p^-}$
are the eigenstates of light-cone momenta
and we assume that the mass-shell condition is satisfied;
\begin{eqnarray}
\left(\al' \frac{p}{R_-}p^- + H^{(\msc{c})} \right)
\ket{p^+=p/R_-}
\otimes \ket{p^-(p,h)}
\otimes \ket{\mbox{D'p},h;\hm p}^{(\msc{tr})}
=0~,
\end{eqnarray}
which uniquely determines $p^-$ as a function of $p$, $h$.

The transverse boundary state 
$\ket{\mbox{D'p};\hm p}^{(\msc{tr})}$ with the mass parameter
$m=\hm p$ is given in \cite{BP}. 
In the present case we have
$\mbp+1$ NN and $7-\mbp$ DD open strings.
Therefore, because of the consistency with \eqn{N w r},
we should choose the normalization coefficients 
$\cN'_p$ as 
\begin{eqnarray}
&& \cN_p' = \frac{1}{\sqrt{2\nu}}\cdot 
\left(2\sinh(\pi \hm p)\right)^{\frac{3-\msp}{2}}~, 
\end{eqnarray}
as presented in \cite{BGG}.
The factor $\dsp \frac{1}{\sqrt{2\nu}}$ is the characteristic one for 
the Dirichlet boundary states which reflects the absence of 
zero-mode integral in the calculation of open string channel.

The free energy \eqn{free energy bs}
can be interpreted as the trace over a subspace of the physical
Hilbert space of closed string sector. To be more accurate this subspace
is composed of all the single-particle physical states that can be
emitted/absorbed by the D'p-brane. We thus regard \eqn{free energy bs}
as the free energy for the thermal ensemble of such closed string states.

It is a straightforward calculation to evaluate \eqn{free energy bs}
and we obtain
\begin{eqnarray}
&& F(\beta;\mbox{D'p}) =- \frac{1}{\beta}\sum_{p=1}^{\infty}\,\left\lb 
\sum_{n\,:\,\msc{even}, n>0}\,\frac{1}{n}e^{-\frac{\beta^2n^2}{2\pi \al' T}}
 \cdot \frac{1}{2\nu}\left(2 \sinh (\pi \hm p)\right)^{3-\msp} \right. \nn 
&& \left. \hspace{1in} + \sum_{n\,:\,\msc{odd}, n>0}\,
\frac{1}{n}e^{-\frac{\beta^2n^2}{2\pi \al' T}}\cdot \frac{1}{2\nu}
\left(2 \sinh (\pi \hm p)\right)^{3-\msp}\cdot 
\frac{\th_{(0,\frac{1}{2})}(T;\hm p)^4}{\th_{(0,0)}(T;\hm p)^4}
\right\rb~,
\label{free energy D'p 1}
\end{eqnarray}
where we set $\dsp T= \frac{2n\nu}{p}$.
Furthermore, we can rewrite it by 
a short calculation based on the modular property 
\eqn{modular th} as follows;
\begin{eqnarray}
&& F(\beta;\mbox{D'p}) =- \frac{1}{\beta}\sum_{\ep=0,1}\,\sum_{q=1}^{\infty}\,
\sum_{w \in 2\bsz+\ep}\, \int_0^{\infty}dt\,
e^{-\frac{\beta^2 w^2 t}{2\pi \al'}} \, \delta(2w\nu t -q) \nn
&& \hspace{1in} \times \left(2 \sinh(2 \pi \hm w\nu t)\right)^{3-\msp}
\cdot \frac{\th_{(\frac{\ep}{2},0)}(t;2\hm w\nu)^4}
{\th_{(0,0)}(t;2\hm w \nu)^4}~.
\label{free energy D'p 2}
\end{eqnarray}
This expression has a nice interpretation as the 
virtual open string amplitude possessing the non-vanishing thermal windings.
In fact, the longitudinal part is calculated by summing
over the instantons associated to  the boundary conditions
\begin{eqnarray}
&&X^+(\tau_E, \sigma=\pi) = 
X^+(\tau_E, \sigma=0)+ i\frac{1}{\sqrt{2}}\beta w ~, \nn
&&X^-(\tau_E, \sigma=\pi) = 
X^-(\tau_E, \sigma=0)- i\frac{1}{\sqrt{2}}\beta w + 2\pi R_- r~, \nn
&& \partial_{\tau_E} X^+|_{\sigma=0,\pi} 
= \partial_{\tau_E} X^-|_{\sigma=0,\pi}=0~.
\label{instanton 4}
\end{eqnarray}
The instanton action is equal to 
\begin{eqnarray}
S_{\msc{inst}}(w,r)= \frac{\beta^2w^2t}{2\pi \al'} + 4\pi i t\nu w r~,
\end{eqnarray}
leading to the longitudinal amplitude
\begin{eqnarray}
\sum_{w,r}\,e^{-S_{\msc{inst}}(w,r)}= \sum_{w,q}\, 
e^{-\frac{\beta^2w^2t}{2\pi \al'}}\,\delta(2\nu w t-q)~.
\label{vo longitudinal}
\end{eqnarray}
Notice that we now do not have the zero-mode integral
because of the Dirichlet condition. 

The mass parameter for the transverse sector is evaluated 
from this  instanton configuration \eqn{instanton 4} as  
$m=2\hm \nu w$. We thus obtain the transverse amplitudes 
by the calculations similar to \eqn{bosonic open amplitude 1} and 
\eqn{fermionic open amplitude 1} 
\begin{eqnarray}
\tr_{\cH_w}\left\lb(-1)^{\bsF}\, e^{-2\pi t H^{\msc{(o)}}}\right\rb
&=& \left(2 \sinh(2 \pi \hm w\nu t)\right)^{3-\msp}
\cdot \frac{\th_{(0,0)}(t;2\hm w\nu)^4}
{\th_{(0,0)}(t;2\hm w \nu)^4} 
\equiv
\left(2 \sinh(2 \pi \hm w\nu t)\right)^{3-\msp}~,\nn
&& \hspace{7.5cm} (w\in 2\bz) ~, \nn
\tr_{\cH_w}\left\lb(-1)^{\bsF}\, e^{-2\pi t H^{\msc{(o)}}}\right\rb
&=& \left(2 \sinh(2 \pi \hm w\nu t)\right)^{3-\msp}
\cdot \frac{\th_{(\frac{1}{2},0)}(t;2\hm w\nu)^4}
{\th_{(0,0)}(t;2\hm w \nu)^4}~~~(w\in 2\bz+1)~,
\label{vo transverse}
\end{eqnarray}
where $\cH_w$ denotes the Hilbert space of virtual open string states
with the thermal  winding number $w$.

These results \eqn{vo longitudinal}, \eqn{vo transverse}
correctly reproduce  the free energy \eqn{free energy D'p 2}.
Only the non-trivial difference from the standard cylinder amplitude 
is that the modulus integral is now given as $\dsp \int dt$
instead of $\dsp \int \frac{dt}{t}$. It amounts to that we are now 
calculating $\sim \tr \Delta$ rather than $\sim \tr \ln \Delta$, where
$\Delta$ denotes the open string propagator.

The Hagedorn temperature is likewise  evaluated by observing 
the behavior of virtual open string with $w=1$;
\begin{eqnarray}
F(\beta;\mbox{D'p})~:~\mbox{finite}~
\Longleftrightarrow~ \beta > \beta_H~
\end{eqnarray}
with 
\begin{eqnarray}
\frac{\beta^2_H}{4\pi^2\al'}-4\left(\Delta
\left(\frac{\sqrt{2}\mu\beta_H}{2\pi};\frac{1}{2}\right)
-\Delta\left(\frac{\sqrt{2}\mu\beta_H;0}{2\pi}\right)
\right)  =0~.
\label{H temperature 3}
\end{eqnarray}
$\beta_H$ is again independent of the value $\mbp$.
Moreover, it is easy to see that $T_H\equiv \beta_H^{-1}$ 
evaluated by the equation 
\eqn{H temperature 3} is strictly higher than those for 
\eqn{H temperature 1}, \eqn{H temperature 2}.
Therefore, we conclude that the existence of Euclidean Dp-branes do not 
affect the Hagedorn behavior.

~

\section{Thermal Partition Function of DLCQ Superstring on 6-dimensional
PP-Wave}

\subsection{DLCQ PP-Wave as the Penrose limit of 
Orbifolded $AdS_3 \times S^3$}
\indent 

Before analyzing the thermal partition function, 
let us illustrate  how the 6-dimensional DLCQ pp-wave with 
enhanced SUSY can be derived 
from the orbifolded $AdS_3 \times S^3$. This is almost parallel
to the 10-dimensional argument \cite{MRV}.

We begin with the familiar system of $Q_5$ D5 and $Q_1$ D1, 
or its transforms by the $SL(2;\bz)$-duality more generally. 
The 5-branes are wrapped along the 4-dimensional compact space 
$M^4(\equiv T^4 ~ \mbox{or}~ K3)$. 
The near horizon geometry is described by the background 
$AdS_3\times S^3 (\times M^4)$ parametrized as
\begin{eqnarray}
ds^2=R^2\left\lb -\cosh^2 \rho\, dt^2 + d\rho^2 + \sinh^2 \rho \, d\phi^2
+ d\al^2 + \sin^2 \al\, d\theta^2 + \cos^2 \al\, d\chi^2\right\rb~,
\label{AdS3 S3}
\end{eqnarray}
with the enhanced SUSY (16 supercharges in the sense of
6-dimension\footnote
   {In the sense of 10-dimensional theory we
    have the 24 supercharges for $M^4=T^4$, 
   and 16 supercharges for $M^4=K3$.}).

Let us consider the $\bz_{N}$-orbifoldization along the 4-dimensional space 
transverse to the whole branes, parametrized by complex coordinates
$z_1$, $z_2$. $S^3$ in the above near horizon geometry is 
described as $|z_1|^2+|z_2|^2=R^2$, and hence the relevant 
geometry  is  $AdS_3\times S^3/\bz_N$. 
We also assume the relation of these coordinates as 
\begin{eqnarray}
z_1 = R \sin \al \, e^{i\theta}~,~~~ z_2 = R\cos \al \, e^{i\chi}~.
\end{eqnarray}
It is convenient to recombine  $z_1$, $z_2$ to a single matrix 
\begin{equation}
Z \df \frac{1}{R}
\left(
\begin{array}{cc}
 z_1 & z_2\\
 -\bar{z_2} & \bar{z_1}
\end{array}
\right)~,
\end{equation}
and the isometry of $S^3$ is expressed as $Z\, \mapsto\, g_LZg_R^{-1}$ 
(${}^{\forall}(g_L,g_R)\in SU(2)_L\times SU(2)_R$), 
which corresponds to the R-symmetry group.
We have several choices of $\bz_N$ action summarized as:
\begin{enumerate}
\item The diagonal action $\bz_N \subset SU(2)_D \subset 
SU(2)_L\times SU(2)_R$, namely,
\begin{eqnarray}
&& Z~\longmapsto~e^{\pi in\sigma_3/N}Ze^{-\pi in\sigma_3/N}~,~~~\nn
&& \Longleftrightarrow~~~ z_1 ~\mapsto~z_1~,~~~
z_2 ~\mapsto~e^{2\pi i n/N}z_2~,
\label{non SUSY orbifold}
\end{eqnarray}
where $\dsp 
\sigma_3=
\left(
\begin{array}{cc}
 1&0 \\
 0&-1
\end{array}
\right)$ is the Pauli matrix.
It breaks the full R-symmetry $SU(2)_L\times SU(2)_R$.
we hence obtain no SUSY theory.
\item The left action $\bz_N \subset SU(2)_L$, namely,
\begin{eqnarray}
&& Z~\longmapsto~e^{2\pi in\sigma_3/N}Z~,~~~\nn
&& \Longleftrightarrow~~~ z_1 ~\mapsto~e^{2\pi i n/N}z_1~,~~~
z_2 ~\mapsto~e^{2\pi i n/N}z_2~.
\label{half SUSY orbifold}
\end{eqnarray}
This orbifoldization breaks the $SU(2)_L$ R-symmetry,
leaving the half SUSY (8 supercharges as the 6-dimensional theory).

\end{enumerate}

Now, let us consider the following two types of the Penrose limits:
\begin{description}
 \item[(a)] We choose the null-geodesic located along $\al=0$. Namely,
we set $r=\rho R$, $y=\al R$, $\dsp x^+= \frac{1}{2}(t+\chi)$,
$\dsp x^-=\frac{R^2}{2}(t-\chi)$,  and take the large $R$ limit 
fixing $r,y,x^+,x^-$ to be finite values.
\item[(b)] We choose the null-geodesic located along $\al=\pi/2$. Namely,
we set $r=\rho R$, $y=(\pi/2-\al) R$, $\dsp x^+= \frac{1}{2}(t+\theta)$,
$\dsp x^-=\frac{R^2}{2}(t-\theta)$,  and take the large $R$ limit 
fixing $r,y,x^+,x^-$ to be finite values.
\end{description}

We have several combinations of the choices of $\bz_N$-action and 
the Penrose limits.  Their aspects are summarized as follows;
\begin{itemize}
\item The combination 1-(a): 

The null-geodesic does not lie along the
fixed point locus ($\al=\pi/2$). We so obtain the smooth 6-dimensional
pp-wave;
\begin{eqnarray}
ds^2 = -4dx^+dx^--(r^2+y^2)(dx^+)^2+dr^2 +r^2d\phi^2 +dy^2+y^2d\theta^2~,
\label{6 dim pp}
\end{eqnarray}
which is compatible with the enhanced SUSY
(equal number of supercharges to $AdS_3\times S^3$). 
A similar SUSY enhancement in the non-SUSY orbifold 
is also pointed out in \cite{Takayanagi} for the 10-dimensional pp-wave. 
We can generally turn on the RR-flux and NSNS-flux at the same time,
\begin{eqnarray}
F^{\sR\sR}_{+12}=F^{\sR\sR}_{+34}\sim \mu~,~~~
F^{\sNS}_{+12}=F^{\sNS}_{+34}\sim \gamma~,
\label{6 dim flux}
\end{eqnarray}
depending on the brane charges we set up.

The orbifoldization  \eqn{non SUSY orbifold} acts on the new coordinates
as  
\begin{eqnarray}
x^+~\mapsto~x^+ + \frac{n\pi}{N}~,~~~x^-~\mapsto~x^-+\frac{n\pi R^2}{N}~.
\label{orbifold l c}
\end{eqnarray}
Therefore, under the ``large quiver limit'',  which 
is defined by  taking $R\,\rightarrow\,\infty$ and 
$N\,\rightarrow\,\infty$ limit at the same time with 
the ratio $R^2/N$ fixed to be finite, 
we  obtain the DLCQ pp-wave background; 
\begin{eqnarray}
&& x^+~:~ \mbox{non-compact}~,~~~ x^-\sim x^-+ 2\pi R_-~, \nn
&& R_- \df \frac{R^2}{2N}  ~.
\label{DLCQ pp}
\end{eqnarray}

\item 1-(b):

The null-geodesic lies along the fixed point locus.
We so obtain the non-SUSY pp-wave which has an orbifold singularity 
in the transverse plane.

\item 2-(a) and 2-(b):


In the case 2-(a), the orbifold action amounts to
\begin{eqnarray}
x^+~\mapsto~x^+ + \frac{n\pi}{N}~,~~~x^-~\mapsto~x^-+\frac{n\pi R^2}{N}~,
~~~ \theta~\mapsto~\theta+\frac{2n\pi}{N}~.
\label{orbifold l c 2}
\end{eqnarray}
Therefore, we again obtain the DLCQ pp-wave with 
enhanced SUSY \eqn{DLCQ pp} under the large quiver limit. The case 2-(b) is 
completely parallel.
\end{itemize}

~

We shall concentrate on the DLCQ pp-wave with the enhanced SUSY from now on. 


\subsection{Thermal Partition Function of the 6-dimensional DLCQ pp-wave :
Case of $M^4=T^4$}
\indent 

We first treat the simpler case $M^4=T^4$. 
The GS action in the light-cone gauge  for the background \eqn{6 dim pp}
with the general flux  \eqn{6 dim flux} is given as follows
(\cite{BMN}, see also \cite{RT}); 
\begin{eqnarray}
&& S=\frac{1}{4\pi\al'}\int d^2\sigma\,\left\lb 
|\partial_{\tau} Z_i|^2  - |(\partial_{\sigma}+i\eta)Z_i|^2
-m^2 |Z_i|^2 +\partial_+Y^j\partial_-Y^j \right\rb \nn
&& ~~~  -\frac{i}{2\pi}\int d^2\sigma\,  \left\lb \bar{S}
(\rho^0\partial_{\tau}+\rho^1\partial_{\sigma}+\rho^1\otimes i\eta M
-I\otimes m M) S  \right\rb~,
\label{6 dim action} 
\end{eqnarray}
where we set $m=\al' \mu p^+$, $\eta=\al' \gamma p^+$.
$Z_1\equiv X^1+iX^2$, $Z_2\equiv X^3+iX^4$ are the coordinates 
along the transverse plane in the pp-wave and $Y^j$ are the coordinates 
of $T^4$. 
$\rho^0$, $\rho^1$ denote the 2-dimensional gamma matrices 
for the world-sheet, explicitly defined by
\begin{eqnarray}
\rho^0 =
\left(
\begin{array}{rr}
 0&-1 \\
 1&0
\end{array}
\right) ~,~~~
\rho^1 =
\left(
\begin{array}{rr}
 0&1 \\
 1&0
\end{array}
\right)~,
\end{eqnarray}
and the fermionic coordinates are expressed as 
\begin{eqnarray}
S= 
\left(
\begin{array}{l}
 S^a \\
 \tilde{S}^a
\end{array}
\right)\in {\bf 8_s}\times {\bf 8_s}~, 
~~~ \bar{S}= S^T \rho^0 \equiv 
(\tilde{S}^a,-S^a)~.
\end{eqnarray}
The $8\times 8$ matrix $M$ is defined in terms of the gamma matrices 
$\gamma^I_{a\db}$, $\bar{\gamma}^I_{\da b}$ introduced before as
follows;
\begin{eqnarray}
M\df  \frac{i}{2}(\gamma^1\bar{\gamma}^2+\gamma^3\bar{\gamma}^4)
\equiv i \gamma^1\bar{\gamma}^2\frac{1}{2}(1-\Pi)~.
\label{M}
\end{eqnarray}
Therefore, we can classify the components of GS fermions as 
\begin{itemize}
 \item $\Pi=+1$ : The eigenvalues of $M$ are all zero. We have 
       8 components of the massless and untwisted GS fermions, 
       which we express as $S^{(0)\, a_0}$, 
       $\tilde{S}^{(0)\, a_0}$ ($a_0=1,\ldots, 4$).
 \item $\Pi=-1$ : The eigenvalues of $M$ are $+1$ and $-1$. 
       They correspond to the 8 components of massive and twisted GS 
       fermions. We denote the 4 components with $M=+1$ ($M=-1$)
       as $S^{(+)\, a_+}$, $\tilde{S}^{(+)\, a_+}$ ($a_+=1,2$)
       ($S^{(-)\, a_-}$, $\tilde{S}^{(-)\, a_-}$  ($a_-=1,2$)).  
       The mass and twist parameter are given by $m$, $\pm \eta$
       respectively. 
\end{itemize}

The equation of motion of bosonic coordinates $Z_i$ is given by
\begin{eqnarray}
\partial_{\tau}^2Z_i-(\partial_{\sigma}+i\eta)^2Z_i + m^2 Z_i=0~.
\label{eom 6dim pp}
\end{eqnarray}
Making use of the simple redefinition of variable; 
$\hat{Z}_i \df e^{i\eta \sigma}Z_i$, this equation reduces to 
the simpler Klein-Gordon equation
\begin{eqnarray}
\partial_{\tau}^2\hat{Z}_i-\partial_{\sigma}^2\hat{Z}_i + m^2 \hat{Z}_i=0~.
\label{eom 6dim pp 2}
\end{eqnarray}
$\hat{Z}_i$ are the free massive complex bosons with the twisted 
boundary condition
\begin{eqnarray}
\hat{Z}_i(\tau, \sigma+2\pi) = e^{2\pi i\eta}\hat{Z}_i(\tau,\sigma)~,
\end{eqnarray}
since the original variable $Z_i$ should be single-valued.

The canonical quantization is most easily defined with respect to 
the free fields $\hat{Z}_i$ (and their fermionic counter parts
$\hat{S}^{(+)\,a_+}$ etc.), and 
the world-sheet Hamiltonian is calculated as
\begin{eqnarray}
H = \sum_{n\in \bsz} \left\lb \sqrt{m^2+(n+\eta)^2} N_n^{(+)}
   + \sqrt{m^2+(n-\eta)^2} N_n^{(-)}
   +|n| N_n^{(0)}\right\rb - a(p^+)~,
\label{H 6 dim}
\end{eqnarray}
where $N_n^{(+)}$, $N_n^{(-)}$
denotes the mode counting operators associated 
to the Fourier modes $e^{\pm i (n+\eta)\sigma}$, $e^{\pm i
(n-\eta)\sigma}$ respectively,
and $N_n^{(0)}$ is that for the remaining massless fields $Y^j$, 
$S^{(0)\,a_0}$, $\tilde{S}^{(0)\,a_0}$.
For example, the annihilation operators contained in $\hat{Z}_i$
and the creation operators within $\hat{Z}_i^*$ are counted by
$N_n^{(+)}$.
(We again employ the convention such that the positive and 
negative modes correspond to the left and right-movers respectively,
and the zero-modes are suitably defined so as to diagonalize the Hamiltonian.)
$a(p^+)$ again denotes the normal order constant which is evaluated 
as before.

The DLCQ compactification $X^-\sim X^- + 2\pi R_-$ leads to
the momentum quantization
\begin{equation}
p^+ = \frac{p}{R_-}~, ~~(p\in \bz_{>0})~,
\end{equation}
and the level matching condition
\begin{eqnarray}
P\equiv \sum_n n \left(N^{(+)}_n +N^{(-)}_n + N^{(0)}_n \right) = pk~,
~~~(k \in \bz)~,
\label{level matching 6 dim}
\end{eqnarray}
where $P$ denotes the world-sheet momentum operator {\em associated 
to the original string coordinates $Z_i$, $S^a$, and so on}.
The momentum associated to the hatted fields is also useful,
which is defined as 
\begin{eqnarray}
\hat{P}&=&
\sum_{n\in \bsz} (n+\eta)N^{(+)}_n +\sum_{n\in \bsz} (n-\eta)N^{(-)}_n
+ \sum_{n\in \bsz} n N^{(0)}_n  \nn
& \equiv & P + \eta \hat{h}~,
\label{hat P}
\end{eqnarray}
where $\dsp \hat{h}\equiv \sum_{n\in
\bsz}\left(N^{(+)}_n-N^{(-)}_n\right)$ is the helicity operator
along the transverse plane of pp-wave.
It is convenient to further introduce the notations 
$\hat{\mu}\equiv  \al' \mu/R_-$, 
$\hat{\gamma}\equiv  \al' \gamma/R_-$,
resulting $m= \hm p$, $\eta= \hg p$.

According to the similar arguments for the 10-dimensional 
pp-wave, we can derive the modular invariant expression of thermal 
partition function by the path-integral calculation.
The non-trivial difference is only the existence of twisted boundary 
conditions for the hatted fields. 
As before, we take the instanton gauge $X^+=X^+_{w,n,r,s}$,
with working on the rotated coordinates $z'=e^{i\th_{w,n}}z$, 
$\bar{z}'=e^{-i\th_{w,n}}\bar{z}$ defined 
by \eqn{rotation}. Thanks to \eqn{instanton gauge}, the world-sheet 
action of the transverse coordinates 
has a quadratic form as \eqn{6 dim action}. Especially, 
the bosonic part is written as (on the Euclidean world-sheet)
\begin{eqnarray}
S=\frac{1}{8\pi \al'}\int d^2z'\, 
\left\{(2\partial_{z'}-i\eta')Z_i(2\partial_{\bar{z}'}+i\eta')Z_i^* 
+(2\partial_{z'}+i\eta')Z^*_i(2\partial_{\bar{z}'}-i\eta')Z_i + 2 m'^2|Z_i|^2
\right\}~,
\label{6 dim action 2}
\end{eqnarray}
where the mass parameter $m'$ is equal to 
$\dsp \hm \frac{\nu}{\tau_2}|w\tau-n|$ and the twist parameter $\eta'$
is similarly calculated as $\dsp \eta' = -\hg\frac{\nu}{\tau_2}|w\tau-n|$.
Introducing the field redefinitions 
$\hat{Z}_i=Z_i e^{-i\frac{\eta'}{2}(z'+\bar{z}')}$,
$\hat{Z}^*_i=Z^*_i e^{i\frac{\eta'}{2}(z'+\bar{z}')}$, 
the action \eqn{6 dim action 2} reduces to a simpler form
\begin{eqnarray}
S=\frac{1}{4\pi \al'} \int d^2z'\, 
\left\{2(\partial_{z'}\hat{Z_i}\partial_{\bar{z}'}\hat{Z_i}^*+
\partial_{\bar{z}'}\hat{Z_i}\partial_{z'}\hat{Z_i}^*)  + m'^2|Z_i|^2\right\}~.
\label{6 dim action 3}
\end{eqnarray}
To perform the quantization we must go back to the original coordinates
$z$, $\bar{z}$. The obtained action also has the same form,
since the action \eqn{6 dim action 3} is rotationally 
invariant. However, {\em  the free fields 
$\hat{Z}_i$, $\hat{Z}_i^*$ have non-trivial boundary
conditions\/}, because of the relations
\begin{eqnarray}
&& \hat{Z}_i= Z_i e^{-i\frac{\eta'}{2}(z'+\bar{z}')}\equiv
Z_i e^{-i\frac{\eta'}{2}(e^{i\th_{w,n}}z+e^{-i\th_{w,n}}\bar{z})}~, \nn
&&
\hat{Z}^*_i= Z^*_i e^{i\frac{\eta'}{2}(z'+\bar{z}')}\equiv
Z^*_i e^{i\frac{\eta'}{2}(e^{i\th_{w,n}}z+e^{-i\th_{w,n}}\bar{z})}~.
\end{eqnarray}
We find after a short calculation
\begin{eqnarray}
\hat{Z}_i(z+2\pi, \bar{z}+2\pi)
&=&e^{2\pi i \hg \frac{\nu(-\tau_1 w+n)}{\tau_2}} 
\,\hat{Z}_i(z, \bar{z})~,\nn
\hat{Z}_i(z+2\pi\tau, \bar{z}+2\pi\bar{\tau})
&=&e^{2\pi i \hg \frac{\nu(-|\tau|^2 w+n\tau_1)}{\tau_2}} 
\,\hat{Z}_i(z, \bar{z})~.
\label{hat Z bc}
\end{eqnarray}

In this way we can calculate the desired thermal partition function
as the form  reminiscent of \eqn{part pp 0};
\begin{eqnarray}
Z_{\msc{torus}}(\beta)&=& \int_{\cF}\frac{d^2\tau}{\tau_2^2}\,\nu 
\sum_{\ep_i=0,1} \,\sum_{\stackrel{w\in 2\bsz+\ep_1}{n\in 2\bsz+\ep_2}}
\,\sum_{r,s}\,e^{-S_{\msc{inst}}(w,n,r,s)}\,
Z^{\msc{tr}}_{\ep_1,\ep_2}\left(\tau,\bar{\tau};
m_{w,n}, \eta_{w,n}^{1}, \eta^2_{w,n}\right) ~,
\label{part pp 6 dim 0}
\end{eqnarray}
where the instanton action $S_{\msc{inst}}$ is defined 
in \eqn{inst action 1}. The transverse partition function 
$Z^{\msc{tr}}_{\ep_1,\ep_2}$ depends on the mass parameter
\begin{eqnarray}
m_{w,n} = \hm \frac{\nu}{\tau_2}|w\tau-n|~,
\end{eqnarray}
and also the spatial and temporal twist parameters
\begin{eqnarray}
\eta^1_{w,n} = \hg \frac{\nu(\tau_1w-n)}{\tau_2}~,~~~
\eta^2_{w,n} = \hg \frac{\nu(-|\tau|^2w+n\tau_1)}{\tau_2}~.
\end{eqnarray}
The subscripts $\ep_1$, $\ep_2$ again 
indicate the thermal boundary conditions
of the GS fermions. 
As before, we can readily carry out the summation over $r$, $s$,
since $Z^{\msc{tr}}_{\ep_1,\ep_2}$ only depends on the thermal
windings $w$, $n$. We thus obtain the same periodic delta function 
\begin{eqnarray}
\sim \tau_2\sum_{w,n,p,q}\, \delta^{(2)}((w\nu+ip)\tau-(n\nu+iq)) ~,
\end{eqnarray}
which imposes the constraints \eqn{delta constraints}.
Based on this fact we can replace the parameters $m_{w,n}$, $\eta^1_{w,n}$ and
$\eta^2_{w,n}$ with the simpler ones;
\begin{eqnarray}
m_{w,n}~\rightarrow~ \hm |w\nu+ip|~, ~~~ 
\eta^1_{w,n}~\rightarrow~\hg p~, ~~~ 
\eta^2_{w,n}~\rightarrow~-\hg q~.
\end{eqnarray}
Notice that the (spatial) twist parameter depends on the light-cone 
momentum $p^+=p/R_-$ as $\hg p$ rather than $\hg |w\nu+ip|$. 
This fact is consistent with the existence of spectral flow symmetry 
in the purely NSNS case $\mu=0$, as we will later discuss. 

$Z^{\msc{tr}}_{\ep_1,\ep_2}$ is again expressed 
by means of the massive theta functions 
$\Theta_{(a,b)}(\tau,\bar{\tau};m)$. 
(Recall \eqn{part m t boson}, for example.)
We finally obtain 
\begin{eqnarray}
&& Z_{\msc{torus}}(\beta) = \int_{\cF} \frac{d^2\tau}{\tau_2^2}\,
\nu \sum_{\ep_i=0,1}\sum_{\stackrel{w\in 2\bsz+\ep_1}{n\in 2\bsz+\ep_2}} \,
e^{-\frac{\beta^2|w\tau-n|^2}{4\pi \al' \tau_2}} \tau_2
\delta^{(2)}\left((w\nu+ip)\tau-(n\nu+iq)\right)  \nn
&& \hspace{1in} \times Z^{(0)}_{T^4}(\tau,\bar{\tau}) 
\frac{1}{(4\pi^2\al'\tau_2)^2}
\frac{1}{|\eta(\tau)|^8} \cdot e^{-\pi \tau_2 \ep_1}
\left|\frac{\theta_1\left(\tau, \frac{\ep_1}{2}\tau+\frac{\ep_2}{2}\right)}
{\eta(\tau)}\right|^4 \nn
&& \hspace{1in} 
\times \frac{\Theta_{(\hg p+\frac{1}{2}\ep_1, -\hg q+\frac{1}{2}\ep_2)}
(\tau,\bar{\tau} ; \hm |w\nu+ip|)^2}
{\Theta_{(\hg p, -\hg q)}
(\tau,\bar{\tau} ; \hm |w\nu+ip|)^2}~.
\label{part pp T4 1}
\end{eqnarray}
where 
$Z_{T^4}^{(0)}$ denotes the (bosonic) 
zero-mode part of partition function
of $T^4$. For example, for the rectangular torus with 
the radii $R_1$, $R_2$, $R_3$, $R_4$, it is calculated as
\begin{eqnarray}
Z^{(0)}_{T^4}(\tau,\bar{\tau})= \prod_{i=1}^4
\left\lb 2\pi R_i \sum_{w_i,n_i}\, 
e^{-\frac{\pi R_i^2|w_i\tau-n_i|^2}{\al'\tau_2}}\right\rb~.
\end{eqnarray}
As in the previous analysis,
we can rewrite \eqn{part pp T4 1} as a simpler form 
by setting $w=0$ based on the modular invariance.
A short calculation gives us
\begin{eqnarray}
&&Z_{\msc{torus}}(\beta) = \sum_{p,q} \sum_{n\,:\,\msc{odd}}\,
\frac{1}{n p} e^{-\frac{\beta^2 n^2}{4\pi \al' \tau_2}}\,
Z^{(0)}_{T^4}(\tau,\bar{\tau})\frac{1}{(4\pi^2\al'\tau_2)^2}
\frac{1}{|\eta(\tau)|^8}\cdot \left|\frac{\th_2(\tau)}{\eta(\tau)}\right|^4
\nn
&& \hspace{1in}
\times  \frac{\Theta_{(\hg p,-\hg q+\frac{1}{2})}
(\tau,\bar{\tau};\hat{\mu}p)^2}
{\Theta_{(\hg p,-\hg q)}(\tau,\bar{\tau};\hat{\mu}p)^2}~,
\label{part pp T4 2}
\end{eqnarray}
where we set $\dsp \tau= \frac{q+in\nu}{p}$ and the summation of
$p$, $q$ and $n$ run over the range such that $\tau \in \cS$. 
This is again shown to be identical to the free energy 
of space-time theory \eqn{free energy} (up to the factor $-\beta$)
calculated by the operator formalism.
The check of level matching condition is only the non-trivial task.
In fact, in the expression of \eqn{part pp T4 2}  
the summation over $q$ acts  as  the projection operator
\begin{equation}
\frac{1}{p}\sum_{q} e^{2\pi i \frac{q}{p}\hat{P}-2\pi i \hg q \hat{h}}
\equiv \frac{1}{p} \sum_q e^{2\pi i \frac{q}{p}P}~,
\end{equation}
which ensures the correct level matching condition \eqn{level matching 6 dim}.

Comparing it  with \eqn{part pp 2}, we notice  the absence of the 
``topological term'' including the sum over even $n$.
In fact, the physical spectrum includes 
the same number of bosonic and fermionic BPS states (see, for example,
\cite{HS1}), resulting the vanishing Witten index. This feature 
of course reflects the fact that $T^4$ has the vanishing Euler number
from the view points of dual conformal theory associated to $Sym^N(T^4)$.

The evaluation of Hagedorn temperature is similarly carried out.
We only have to observe the IR behavior of the term with 
$w=1$, $p=n=0$. The result is 
\begin{eqnarray}
Z_{\msc{torus}}(\beta)~:~\mbox{finite}~
\Longleftrightarrow~ \beta > \beta_H~
\end{eqnarray}
with 
\begin{eqnarray}
\frac{\beta^2_H}{8\pi^2\al'}-4\left(\Delta
\left(\frac{\sqrt{2}\mu\beta_H}{4\pi};\frac{1}{2}\right)
-\Delta\left(\frac{\sqrt{2}\mu\beta_H}{4\pi};0\right)
\right)  -\frac{1}{2} =0~.
\end{eqnarray}
Notice that the Hagedorn temperature only depends on the RR-flux $\mu$
and does not on the NSNS one $\gamma$. Especially, among the 
$SL(2;\bz)$-family of $(\mu,\gamma)$, 
the purely NSNS case ($\mu=0$) 
has the minimal Hagedorn temperature (maximal $\beta_H$) which is 
equal to that for the flat-background.

We finally make a comment on the purely NSNS case.
If $\hg$ is an irrational value, nothing interesting happens.
However, in the cases of rational $\hg$,
we gain a periodicity under  $\hg p \, \rightarrow \hg p + r$,  
$\hg q\, \rightarrow \hg q + s$  $({}^{\forall}r, s\in \bz)$, 
and also find a ``resonance'' at the each point 
of $\hg p , \hg q \in \bz$.
In fact, the twisting disappears at these points.
We thus find new zero-modes which arise  a divergent 
volume factor; 
\begin{eqnarray}
\Theta_{(\hg p, - \hg q)}(\tau, \bar{\tau};0)^{-2} = 
\left|\frac{\eta(\tau)}{\th_1(\tau,\hg p\tau - \hg q)}\right|^4
\sim V_4 \times \frac{1}{|\eta(\tau)|^8}~.
\end{eqnarray}
As is already pointed out in \cite{BMN}, 
these extra zero-modes should correspond 
to the ``long strings'' in the original $AdS_3$
background \cite{MMS,SW,MO}, and also the periodicity mentioned above
reflects the spectral flow symmetry.
In the covariant gauge quantization of this background
(i.e. the $H_6$ super WZW model), the long string modes correspond to 
the ``spectrally flowed type I representations'', which describe
the strings freely propagating along the transverse plane,
as is discussed in \cite{HS1} (see also \cite{KK,KP}).


~

\subsection{Case of $M^4=K3$ : Orbifold Point}
\indent

We next analyze  the more non-trivial case $M^4=K3$.
We shall take the simplest orbifold point in the $K3$ moduli space,
namely, the $\bz_2$-orbifold of $T^4$. 

The symmetry group acting on the tangent space is 
\begin{equation}
SO(2)\times SO(2) \times SO(4)_{T^4} \sim 
U(1)\times U(1) \times \widetilde{SU}(2)_L\times \widetilde{SU}(2)_R~,
\end{equation}
and the string coordinates are classified by the representations
of this group as follows;
\begin{eqnarray}
&&Z_1~ (Z_1^*)~:~(\B{1},\B{1})^{+1,0}~((\B{1},\B{1})^{-1,0})~,~~~
Z_2~ (Z_2^*)~:~(\B{1},\B{1})^{0,+1}~((\B{1},\B{1})^{0,-1})~,~~~\nn
&& Y^j~:~(\B{2},\B{2})^{0,0}~, \\
&& S^{(\pm)\,a_{\pm}},~ \tilde{S}^{(\pm)\,a_{\pm}} ~
:~(\B{1},\B{2})^{\pm \frac{1}{2},\pm \frac{1}{2}}~,~~~
S^{(0)\,a_0},~ \tilde{S}^{(0)\,a_0}~:~
(\B{2},\B{1})^{\pm \frac{1}{2}, \mp \frac{1}{2}}
\oplus (\B{2},\B{1})^{\mp \frac{1}{2},\pm \frac{1}{2}}~,
\end{eqnarray}
where the superscripts indicate the $U(1)\times U(1)$-charges
along the pp-wave directions.
We assume the $\bz_2$ action is a subgroup of $\widetilde{SU}(2)_L$.
Namely, it acts as 
\begin{eqnarray}
Y^j~\mapsto ~ -Y^j~,~~~ S^{(0)\,a_0}~\mapsto~-S^{(0)\,a_0}~,~~~
\tilde{S}^{(0)\,a_0}~\mapsto~-\tilde{S}^{(0)\,a_0}~,
\end{eqnarray} 
and other coordinates remain invariant.
This orbifoldization does not break any SUSY in the sense of 
6-dimension\footnote
    {In the sense of 10-dimensional theory, it of course  breaks 
     the 8 supercharges corresponding to 
     the fermionic zero-modes of $S^{(0)\,a_0}$,
     $\tilde{S}^{(0)\,a_0}$, which are the superpartners of 
     $T^4$-coordinates $Y^j$ with respect to the dynamical
     supercharges. Hence the number of Killing spinors is reduced from 
     24 to 16 by the orbifoldization.}, leaving the 6-dimensional
pp-wave with the enhanced SUSY.

The calculation of partition function is carried out 
based on the standard orbifold procedure.
The result is written as 
\begin{eqnarray}
Z_{\msc{torus}}= Z^{\msc{u}}_{\msc{torus}}+Z^{\msc{t}}_{\msc{torus}}~,
\label{part pp K3 1}
\end{eqnarray}
where the contribution from the untwisted sector
$Z^{\msc{u}}_{\msc{torus}}$ is equal to the half of the partition 
function for the $T^4$ case \eqn{part pp T4 1}.
The partition function for the twisted sectors is given by
\begin{eqnarray}
&&Z^{\msc{t}}_{\msc{torus}}(\beta)=\frac{16}{2}
\int_{\cF}\frac{d^2\tau}{\tau_2}\, \nu 
\sum_{\stackrel{\sigma_i=0,1}{(\sigma_1,\sigma_2)\neq (0,0)}}\,
\sum_{\ep_i=0,1}\,
\sum_{\stackrel{w\in 2\bsz+\ep_1}{n\in 2\bsz +\ep_2}}\,\sum_{p,q}\,
e^{-\frac{\beta^2|w\tau-n|^2}{4\pi \al'\tau_2}} \tau_2
\delta^{(2)}\left((w\nu+ip)\tau-(n\nu+iq)\right)  \nn
&& \hspace{1in}
\times e^{-\pi \tau_2(\ep_1+\sigma_1)^2+\pi \tau_2 \sigma_1}\left|
\frac{\th_1\left(\tau,\frac{\sigma_1+\ep_1}{2}\tau
+\frac{\sigma_2+\ep_2}{2}\right)}
{\th_1\left(\tau,\frac{\sigma_1}{2}\tau
+\frac{\sigma_2}{2}\right)}
\right|^4 \nn
&& \hspace{1in} \times
\frac{\Theta_{(\hg p +\frac{\ep_1}{2}, -\hg q+\frac{\ep_2}{2})}
(\tau,\bar{\tau};\hm |w\nu+ip|)^2}
{\Theta_{(\hg p, - \hg q)}
(\tau,\bar{\tau};\hm |w\nu+ip|)^2}~,
\label{part pp K3 t 1}
\end{eqnarray}
where the numerical factor 16 is due to the number of fixed points.

Let us further rewrite it by setting $w=0$ as before.
The twisted sector \eqn{part pp K3 t 1} now yields the non-vanishing
topological term, which is  evaluated as 
\begin{eqnarray}
\frac{16}{2}\sum_{n\,:\,\msc{even}}\,\sum_{p,q}\, \frac{1}{pn}
e^{-\frac{\beta^2n^2}{4\pi\al'\tau_2}}\cdot
\sum_{(\sigma_1, \sigma_2)\neq (0,0)}1 \equiv
24 \sum_{n\,:\, \msc{even}}\,\sum_{p,q}\, 
\frac{1}{pn}e^{-\frac{\beta^2n^2}{4\pi\al'\tau_2}}~.
\end{eqnarray} 
The numerical factor 24 is indeed equal to the Witten index
and the Euler number of $K3$ as should be. In fact, we can show
that the physical spectrum includes the 24 bosonic BPS states
and no fermionic BPS states for each fixed light-cone momentum 
$p^+\equiv p/R_-$, as is analyzed in \cite{HS1} 
for the purely NSNS case.  
The desired partition function is finally written as  
\begin{eqnarray}
&& Z_{\msc{torus}}(\beta) = 24 \sum_{n\,:\,\msc{even}}\,\sum_{p,q}\,
\frac{1}{pn}e^{-\frac{\beta^2n^2}{4\pi\al'\tau_2}} \nn
&& \hspace{1in} + \frac{1}{2}\sum_{n\,:\,\msc{odd}}\,\sum_{p,q}\,
\frac{1}{pn}e^{-\frac{\beta^2n^2}{4\pi\al'\tau_2}}\, \left\lb 
Z^{(0)}_{T^4}(\tau,\bar{\tau}) \frac{1}{(4\pi^2 \al'\tau_2)^2}
\frac{1}{|\eta(\tau)|^8}\cdot \left|\frac{\th_2(\tau)}{\eta(\tau)}\right|^4
\right. \nn
&& \hspace{1in} \left. + \sum_{(\sigma_1,\sigma_2)\neq (0,0)}\, 
\left|\frac{
\th_2\left(\tau,\frac{\sigma_1}{2}\tau+\frac{\sigma_2}{2}\right)}
{\th_1\left(\tau,\frac{\sigma_1}{2}\tau+\frac{\sigma_2}{2}\right)}
\right|^4  \right\rb \cdot
\frac{\Theta_{(\hg p, -\hg q+\frac{1}{2})}
(\tau,\bar{\tau};\hm p)^2}
{\Theta_{(\hg p, -\hg q)}
(\tau,\bar{\tau};\hm p)^2}~.
\label{part pp K3 2}
\end{eqnarray}

~

\subsection{Case of $M^4=K3$ : General Gepner Points}
\indent

As a consistency check let us consider the general Gepner constructions
\cite{Gepner} of $M^4=K3$.   
We focus on the case of purely NSNS flux and employ the RNS
formalism, since it is  still a difficult problem to work on the models 
with the general flux in this situation.

In \cite{EOTY} the Gepner models for $K3$ is studied in detail.
The general form of the partition function of $K3$ non-linear 
$\sigma$-model is written as 
\begin{eqnarray}
Z_{K3}(\tau,\bar{\tau})
=\frac{1}{2}\sum_{\al}\sum_{I}D_I\left|F^{(\al)}_I(\tau)\right|^2~,
\label{K3 sigma}
\end{eqnarray}
which is defined as the diagonal modular invariant with respect to
the spin structures $\al= \NS,\,\tNS,\,\R,\,\tR$. 
The conformal blocks $F^{(\al)}_I(\tau)$ are constructed 
from the characters of $\cN=2$ minimal models, being summed up 
over the integral spectral flows. 
The coefficients $D_I$ are the positive 
integers characterizing the degeneracies of conformal blocks.
The overall normalization is determined uniquely
so that the ``graviton orbit'' (the conformal block 
including the identity representation) has the degeneracy 1.

We also introduce the functions
\begin{eqnarray}
&& f^{(\sNS)}_{(u,v)}(\tau) \df 
\frac{\th_3(\tau,u\tau+v)}{\th_1(\tau,u\tau+v)}~,~~~
f^{(\stNS)}_{(u,v)}(\tau) \df 
\frac{\th_4(\tau,u\tau+v)}{\th_1(\tau,u\tau+v)}~,~~~\nn
&&
f^{(\sR)}_{(u,v)}(\tau) \df 
\frac{\th_2(\tau,u\tau+v)}{\th_1(\tau,u\tau+v)}~,~~~
f^{(\stR)}_{(u,v)}(\tau) \df 
\frac{\th_1(\tau,u\tau+v)}{\th_1(\tau,u\tau+v)} \equiv 1 ~,
\end{eqnarray}
which are convenient to describe the conformal blocks for the NSNS pp-wave
background in the RNS formalism. 

The evaluation of thermal partition function is almost parallel.
However, we have to make a little modification for the thermal boundary 
condition of fermionic coordinates, since we are now working with
the RNS fermions rather than the GS ones. 
To this aim it is convenient to introduce the 
next phase factors depending on the thermal winding numbers $w$, $n$  
and the spin structures;
\begin{eqnarray}
&&\kappa(\NS;w,n) \df 1~, ~~~\kappa(\tNS;w,n) \df (-1)^w~,~~~ \nn
&&\kappa(\R;w,n) \df (-1)^n~,~~~\kappa(\tR;w,n) \df (-1)^{w+n}~.
\end{eqnarray}
When $w=0$, these phase factors reproduce the correct boundary 
condition for the fermionic particle theory with finite temperature, and 
the $w$-dependence is determined by the consistency with the modular 
invariance \cite{AW}.

Under these preparations we can write down the modular invariant form
of thermal partition function;
\begin{eqnarray}
&&Z_{\msc{torus}}(\beta)= \frac{1}{4}\int_{\cF}\frac{d^2\tau}{\tau_2}\,
\nu \sum_{\al,\bar{\al}}\sum_I\sum_{w,n,p,q}\,
e^{-\frac{\beta^2|w\tau-n|^2}{4\pi\al' \tau_2}}\tau_2
\delta^{(2)}\left((w\nu+ip)\tau-(n\nu+iq)\right) \nn
&& ~~ \times \kappa(\al;w,n)\kappa(\bar{\al};w,n) \cdot
\left(f^{(\al)}_{(\hg p, -\hg q)}(\tau)
f^{(\bar{\al})}_{(\hg p, -\hg q)}(\tau)^* \right)^2 
\cdot \ep(\al)\ep(\bar{\al})D_I F^{(\al)}_I(\tau)F^{(\bar{\al})}_I(\tau)^*~,
\end{eqnarray} 
where $\ep(\al)$ is defined by
\begin{eqnarray}
\ep(\NS)\df +1~, ~~~\ep(\tNS)\df -1~,~~~\ep(\R)\df -1~,~~~\ep(\tR)\df +1~,
\end{eqnarray}
which impose the correct GSO projection.
We can also rewrite it by setting $w=0$ as before.
For this purpose we first notice the following identity
\begin{equation}
\sum_{\al}\ep(\al)f^{(\al)}_{(u,v)}(\tau)f^{(\al)}_{(-u,-v)}(\tau)
F^{(\al)}_I(\tau) \equiv 0 ~, ~~~({}^{\forall}u,v)~.
\label{identity 1}
\end{equation}
This identity generically holds not depending on
the detailed structure of Gepner models.
It is most easily proved by the general theorem
about the character formulas of the ``$c=12$ extended superconformal
algebra'' presented in the appendix B of \cite{HS2} 
(See also \cite{ET,Odake,EOTY}).
We also remark the simple relations
\begin{eqnarray}
&& f^{(\al)}_{(-u,-v)}(\tau)= - f^{(\al)}_{(u,v)}(\tau)~, ~~~
(\al = \NS,\, \tNS,\, \R) ~,\nn
&& f^{(\stR)}_{(-u,-v)}(\tau)=f^{(\stR)}_{(u,v)}(\tau) (\equiv 1)~.
\label{identity 2}
\end{eqnarray}
We thus obtain
\begin{eqnarray}
&& \sum_{\al}\ep(\al)f^{(\al)}_{(u,v)}(\tau)^2 F^{(\al)}_I(\tau)
= 2 f^{(\stR)}_{(u,v)}(\tau)^2 F^{(\stR)}_I(\tau) \equiv
2F^{(\stR)}_I(\tau) ~, 
\label{identity 3}\\
&& \sum_{\al}\ep(\al)\kappa(\al;0,2k+1)
f^{(\al)}_{(u,v)}(\tau)^2 F^{(\al)}_I(\tau)
= 2 f^{(\sR)}_{(u,v)}(\tau)^2 F^{(\sR)}_I(\tau)~. 
\label{identity 4}
\end{eqnarray}
$F^{(\stR)}_I(\tau)$ is no other than the Witten index, and
it is known \cite{EOTY} that 
\begin{equation}
\sum_I D_I |F^{(\stR)}_I(\tau)|^2 = \chi(K3) \equiv 24~,
\label{identity 5}
\end{equation}
irrespective of the choice of Gepner models describing $K3$.

With the helps of the identities \eqn{identity 3}, \eqn{identity 4} and
\eqn{identity 5}, we finally obtain
\begin{eqnarray}
&&Z_{\msc{torus}}(\beta)=24\sum_{n\,:\,\msc{even}}\,\sum_{p,q}\,
\frac{1}{pn}e^{-\frac{\beta^2n^2}{4\pi \al' \tau_2}} \nn
&& \hspace{1in} + \sum_{n\,:\,\msc{odd}}\,\sum_{p,q}\
\frac{1}{pn} e^{-\frac{\beta^2n^2}{4\pi \al' \tau_2}}
\cdot \left|f^{(\sR)}_{(\hg p, -\hg q)}(\tau)\right|^4\cdot \sum_ID_I
\left|F_I^{(\sR)}(\tau)\right|^2~,
\label{part pp K3 3} 
\end{eqnarray}
where the summation is taken over the range 
$\dsp \tau\equiv \frac{q+in\nu}{p}\in \cS$ as before.
The topological term is equal to that in our previous result
\eqn{part pp K3 2}, which implies the consistency of calculation.
The second term is sensitive to the moduli of $K3$. 
Comparing \eqn{part pp K3 3} with \eqn{part pp K3 2},
we conjecture that the  partition function for the general flux 
$(\mu, \gamma)$ and the general Gepner points of $K3$ 
\eqn{K3 sigma} is given by
\begin{eqnarray}
&&Z_{\msc{torus}}(\beta)=24\sum_{n\,:\,\msc{even}}\,\sum_{p,q}\,
\frac{1}{pn}e^{-\frac{\beta^2n^2}{4\pi \al' \tau_2}} \nn
&& \hspace{1in} + \sum_{n\,:\,\msc{odd}}\,\sum_{p,q}\
\frac{1}{pn} e^{-\frac{\beta^2n^2}{4\pi \al' \tau_2}}
\cdot 
\frac{\Theta_{(\hg p, -\hg q+\frac{1}{2})}
(\tau,\bar{\tau};\hm p)^2}
{\Theta_{(\hg p, -\hg q)}
(\tau,\bar{\tau};\hm p)^2}
\cdot \sum_ID_I
\left|F_I^{(\sR)}(\tau)\right|^2~.
\label{part pp K3 4} 
\end{eqnarray}
This might  be proved  by the covariant quantization based on 
the so-called hybrid formalism developed  in \cite{Berkovits}
(see also \cite{BerkM}), although it is beyond the scope of this paper.

~

\section{Summary and Discussions}
\indent

In this paper we have calculated the one-loop thermal amplitudes 
for the closed and open strings on  the DLCQ pp-waves with enhanced 
SUSY.  All these amplitudes can be calculated 
by the operator formalism as the forms that only contain
the contributions from the physical states compatible 
with the standard light-cone gauge. 
However, the path-integral approach is very useful in order to 
derive directly the manifestly modular invariant expressions,
which include the sectors of ``virtual strings'' possessing 
the non-vanishing thermal windings and have the modified mass parameters. 
The virtual strings yield a simple evaluation of Hagedorn temperature,
and further make it possible to achieve the correct open-closed 
string duality for the cylinder amplitudes.

The existence of Hagedorn behavior is not so surprising
and the analysis on it is almost parallel to the flat case,
although the equation determining the Hagedorn temperature 
is affected non-trivially by the  mass deformation. 
However, as a possible direction for future study, it may be 
interesting to explore  the relation to the well-known
thermal phase transition between the thermal $AdS$ 
and the black hole embedded into  the $AdS$ space 
discussed in \cite{Witten,MS}. Especially, in the case of 
$AdS_3 \times S^3$ \cite{MS}, the aspects of  thermal phase transition
between the thermal $AdS_3$ and the BTZ black hole is 
finely controlled by the modular transformation on the 
{\em boundary torus\/}. In this context the inverse temperature $\beta$
should be identified with the modulus $\tau_2$ for the boundary torus,
on which the dual $SCFT_2$ is defined.
On the other hand,  at least for the flat background, 
the {\em space-time\/} modulus $\beta$ in the thermal DLCQ superstring
theory is known to be identified with the {\em world-sheet\/} modulus
for the Matrix string theory \cite{Semenoff}. 
Therefore, our thermodynamical analysis on the DLCQ pp-waves 
would  bring a helpful insight in order to understand the aspects of
these phase transitions at the stringy level.

The thermodynamical analysis for the dual quiver gauge theory
(for the case of 10-dimensional DLCQ pp-wave) given in \cite{MRV}
will be also an important future study. It is interesting
to discuss to what extent we can correctly reproduce 
the Hagedorn temperature based on the perturbative calculation 
in the large quiver gauge theory.  
For the 6-dimensional case, the orbifolded $AdS_3\times S^3$
should be dual to an $\cN=(0,4)$ $SCFT_2$. A quiver formulation of 
the dual $SCFT_2$ based on the symmetric orbifold theory is discussed 
in \cite{Sugawara}.  It may be interesting to investigate the large 
quiver limit of such $SCFT_2$ as the model describing  the 6-dimensional 
DLCQ pp-wave, and perhaps, to discuss the relation with the string bit 
model \cite{Verlinde}.

~

~


\section*{Acknowledgement}
\indent

I would like to thank Y. Hikida and  T. Takayanagi for valuable discussions.
This is supported in part by a Grant-in-Aid for 
the Encouragement of Young Scientists 
($\sharp 13740144$) from the Japanese Ministry of Education, 
Culture, Sports, Science and Technology.

~

~


\section*{Appendix ~ Some Notations}
\setcounter{equation}{0}
\def\theequation{A.\arabic{equation}}

We here summarize the convention of theta functions.
We set $q\equiv e^{2\pi i \tau}$, $y\equiv e^{2\pi i z}$.
\begin{eqnarray}
&& \th_1(\tau,z) =i\sum_{n=-\infty}^{\infty}(-1)^n q^{(n-1/2)^2/2} y^{n-1/2}
  \equiv 2 \sin(\pi z)q^{1/8}\prod_{m=1}^{\infty}
    (1-q^m)(1-yq^m)(1-y^{-1}q^m), \nn
&&  \th_2(\tau,z)=\sum_{n=-\infty}^{\infty} q^{(n-1/2)^2/2} y^{n-1/2}
  \equiv 2 \cos(\pi z)q^{1/8}\prod_{m=1}^{\infty}
    (1-q^m)(1+yq^m)(1+y^{-1}q^m), \nn
&& \th_3(\tau,z)=\sum_{n=-\infty}^{\infty} q^{n^2/2} y^{n}
  \equiv \prod_{m=1}^{\infty}
    (1-q^m)(1+yq^{m-1/2})(1+y^{-1}q^{m-1/2}), \nn
&& \th_4(\tau,z)=\sum_{n=-\infty}^{\infty}(-1)^n q^{n^2/2} y^{n}
  \equiv \prod_{m=1}^{\infty}
    (1-q^m)(1-yq^{m-1/2})(1-y^{-1}q^{m-1/2}) .
\end{eqnarray}
 We also use the standard convention of $\eta$-function;
 \begin{equation}
 \eta(\tau)=q^{1/24}\prod_{n=1}^{\infty}(1-q^n).
 \end{equation}

~

\newpage

\end{document}